\documentclass[preprint,aps,prd,amsmath,amssymb]{revtex4}
\pdfoutput=1
\usepackage{graphicx}
\usepackage{color}
\newcommand{\beq}{\begin{eqnarray}}
\newcommand{\eeq}{\end{eqnarray}}

\newcommand{\bmp}{\noindent\begin{minipage}{16cm}}
\newcommand{\emp}{\end{minipage}\vskip 7mm} % 7mm untightened

\usepackage{graphicx}% Include figure files
\usepackage{dcolumn}% Align table columns on decimal point
\usepackage{bm}% bold math
\usepackage{amsmath}
\usepackage{amsfonts}
\usepackage{bbm}
\usepackage{subfigure}
\usepackage{ulem}

\def\drawbox#1#2{\hrule height#2pt
        \hbox{\vrule width#2pt height#1pt \kern#1pt
              \vrule width#2pt}
              \hrule height#2pt}
\def\Fund#1#2{\vcenter{\vbox{\drawbox{#1}{#2}}}}
\def\Asym#1#2{\vcenter{\vbox{\drawbox{#1}{#2}
              \kern-#2pt % line up boxes
              \drawbox{#1}{#2}}}}

\def\fund{\Fund{6.4}{0.3}}

\begin{document}

%%%%%%%%%%%%%%%%%%%%%%%%%%%%%%%%%%%%%%%%%%%%%%%%%%%%%%%%%%%%%%%%%%%%%%%%%%%
\title{\hfill\vbox{\hbox{\rm\small CERN-PH-TH/2008-188}}\\{\Large Ultra Minimal Technicolor and its Dark Matter TIMP} }
\author{Thomas A. {\sc Ryttov}$^{a,c}$}
\email{ryttov@nbi.dk}
\author{Francesco {\sc Sannino}$^{b}$}
\email{sannino@ifk.sdu.dk} \affiliation{\mbox{$^a$CERN Theory Division,
CH-1211 Geneva 23, Switzerland} \\ \mbox{$^b$ High Energy Center, University of Southern Denmark, Campusvej 55, DK-5230 Odense M} \\
\mbox{$^c$Niels Bohr Institute, Blegdamsvej 17, DK-2100 Copenhagen, Denmark}}
%%%%%%%%%%%%%%%%%%%%%%%%%%%%%%%%%%%%%%%%%%%%%%%%%%%%%%%%%%%%%%%%%%%%%%%%%%%%%%%%%%%%%%%%

\begin{abstract}

We introduce an explicit model with technifermion matter transforming according to multiple representations of the underlying technicolor gauge group. The model features simultaneously the smallest possible value of the naive $S$ parameter and the smallest possible number of technifermions. The chiral dynamics is extremely rich. We construct the low-energy effective Lagrangian. We provide both the linearly and non-linearly realized ones. We then embed, in a natural way, the Standard Model (SM) interactions within the global symmetries of the underlying gauge theory.
Several low-energy composite particles are SM singlets.  One of these Technicolor Interacting Massive Particles (TIMP)s is a natural cold dark matter (DM) candidate. We estimate the fraction of the mass in the universe constituted by our DM candidate over the baryon one. We show that the new TIMP, differently from earlier models, can be sufficiently light to be directly produced and studied at the Large Hadron Collider (LHC). \end{abstract}

%%%%%%%%%%%%%%%%%%%%%%%%%%%%%%%%%%%%%%%%%%%%%%%%%%%%%%%%%%%%%%%%%%%%%%%%
\maketitle
\newpage

\section{Introduction}

Understanding the origin of the electroweak symmetry breaking and its possible relation to DM constitute two of the most profound theoretical challenges at present. New strong dynamics at the electroweak scale  \cite{Weinberg:1979bn,Susskind:1978ms} may very well provide a solution to the problem of the origin of the bright and dark \cite{Nussinov:1985xr,Barr:1990ca,Gudnason:2006yj} mass.  A large class of models has recently been proposed \cite{Sannino:2004qp} which makes use of higher dimensional representations of the underlying technicolor gauge group. This has triggered much work related to both the LHC phenomenology, lattice studies and DM \cite{Foadi:2007ue,Gudnason:2006mk,Dietrich:2005jn,Dietrich:2006cm,Ryttov:2007sr,Catterall:2007yx,
DelDebbio:2008wb,Catterall:2008qk,Evans:2005pu,Foadi:2008ci,Gudnason:2006yj,
Kouvaris:2007iq,Kouvaris:2008hc,Cline:2008hr,Dietrich:2008ni}. For a recent review see \cite{Sannino:2008ha}.

Here we provide an explicit example of (near) conformal (NC) technicolor \cite{Holdom:1984sk,Holdom:1983kw,Eichten:1979ah,Holdom:1981rm,Yamawaki:1985zg,Appelquist:an,Appelquist:1999dq} with two types of technifermions, i.e. transforming according to two different representations of the underlying technicolor gauge group \cite{Dietrich:2006cm,Lane:1989ej}. The model possesses a number of interesting properties to recommend it over the earlier models of dynamical electroweak symmetry breaking:
\begin{itemize}
\item
 Features the lowest possible value of the naive $S$ parameter \cite{Peskin:1990zt,{Peskin:1991sw}} while possessing a dynamics which is NC.

 \item Contains, overall, the lowest possible number of fermions. 

\item Yields natural DM candidates.
\end{itemize}
Due to the above properties we term this model {\it Ultra Minimal near conformal Technicolor} (UMT). It is constituted by an $SU(2)$ technicolor gauge group with two Dirac flavors in the fundamental representation also carrying electroweak charges, as well as, two additional Weyl fermions in the adjoint representation but singlets under the SM gauge groups.

In the next section we arrive at this specific UMT model using the conjectured all-orders beta function for nonsupersymmetric gauge theories \cite{Ryttov:2007cx}. In Section \ref{model} we write the underlying Lagrangian and identify the global symmetries of the theory before and after dynamical symmetry breaking. We then construct both the linearly and non-linearly realized low-energy effective Lagrangians. We naturally embed the Standard Model (SM) interactions within the global symmetries of the underlying gauge theory. Several low-energy composite particles are SM singlets. In particular there is a di-techniquark state which is a possible cold DM candidate. This Technicolor Interacting Massive Particle (TIMP) is a natural cold DM candidate as shown in Section \ref{timp}. We also estimate the fraction of the mass in the universe constituted by our DM candidate over the baryon one as function of the Lepton number and the DM mass. The new TIMP, differently from earlier models \cite{Nussinov:1985xr,Barr:1990ca}, can be sufficiently light to be directly produced and studied at the Large Hadron Collider (LHC).  The expected rate of events detectable in experiments such as CDMS \cite{Ahmed:2008eu}, as function of the DM mass, is computed showing that it is not constrained by current data. We draw our conclusions in the final section.

\section{From the Conformal Window to Ultra Minimal Technicolor}

To construct a realistic model of electroweak symmetry breaking one is faced with the constraints coming from the electroweak precision tests. Specifically the new physics beyond the SM must not give a too large contribution to the $S$ parameter. Consider an $SU(N)$ technicolor theory with $N_f$ Dirac fermions in the representation $r$. The naive estimate of $S$ computed in the approximation of a techniquark loop with momentum-independent constituent masses much heavier than the $Z$ mass \footnote{See also \cite{Kurachi:2006mu,Kurachi:2006ej} for a computation of the $S$ parameter when allowing for a dynamical techniquark mass. This computation suggests, in agreement with the results in \cite{Appelquist:1998xf}, that the $S$ parameter is further reduced with respect to the value from the naive analysis.} is given by
\begin{eqnarray}
S = \frac{1}{6\pi} \frac{N_f}{2}d(r) \ ,
\end{eqnarray}
where $d(r)$ is the dimension of the representation $r$. From the estimate above it is clear that an $SU(2)$ technicolor theory with two Dirac fermions in the fundamental representation yields the smallest possible contribution.

However for this low number of flavors the theory is far from possessing NC dynamics and the naive $S$ value underestimates the physical value \cite{Peskin:1990zt,Peskin:1991sw}. The situation changes for NC theories \cite{Appelquist:1998xf}.

Insisting on a NC model with this minimal $S$ parameter an obvious way to obtain conformality is to add the remaining fundamental flavors, neutral under the electroweak symmetries, needed to be just outside the conformal window. The near conformal technicolor theories constructed in this way have been termed partially gauged technicolor  \cite{Dietrich:2006cm}. However, as we shall show below, by arranging the additional fermions in higher dimensional representations, it is possible to construct models which have a particle content smaller than the one of partially gauged technicolor theories. In fact instead of considering additional fundamental flavors we shall consider adjoint flavors. Note that for two colors there exists only one distinct two-indexed representation.

How many adjoint fermions are needed to build the above NC model? Information on the conformal window for gauge theories containing fermions transforming according to distinct representations is vital. First principle lattice simulations are exploring the conformal window for higher dimensional representations \cite{Catterall:2007yx,DelDebbio:2008wb,Catterall:2008qk}. However the models we are constructing have not yet been explored on the lattice.

To elucidate the various possibilities we make use of our recently conjectured all-order beta function for a generic $SU(N)$ gauge theory with fermionic matter transforming according to arbitrary representations \cite{Ryttov:2007cx}. Considering $N_f(r_i)$ Dirac flavors belonging to the representation $r_i,\ i=1,\ldots,k$ of the gauge group it reads
\begin{eqnarray}
\beta(g) &=&- \frac{g^3}{(4\pi)^2} \frac{\beta_0 - \frac{2}{3}\, \sum_{i=1}^k T(r_i)\,N_{f}(r_i) \,\gamma_i(g^2)}{1- \frac{g^2}{8\pi^2} C_2(G)\left( 1+ \frac{2\beta_0'}{\beta_0} \right)} \ ,
\end{eqnarray}
with
\begin{eqnarray}
\beta_0 =\frac{11}{3}C_2(G)- \frac{4}{3}\sum_{i=1}^k \,T(r_i)N_f(r_i) \qquad \text{and} \qquad \beta_0' = C_2(G) - \sum_{i=1}^k T(r_i)N_f(r_i)  \ .
\end{eqnarray}

One should note that the beta function is given in terms of the anomalous dimension of the fermion mass $\gamma=-\frac{d\ln m}{d\ln \mu}$ where $m$ is the renormalized mass, similar to the supersymmetric case \cite{Novikov:1983uc,Shifman:1986zi,Jones:1983ip}. Indeed the construction of the above beta function is inspired by the one of their supersymmetric cousin theories. At small coupling it coincides with the two-loop beta function and in the non-perturbative regime reproduces earlier known exact results. Similar to the supersymmetric case it allows for a bound of the conformal window \cite{Seiberg:1994pq}. In the supersymmetric case where additional checks can be made the bound is actually believed to give the true conformal window. We stress that the predictions of the conformal window coming from the above beta function are nontrivially supported by all the recent lattice results \cite{Catterall:2007yx,DelDebbio:2008wb,Catterall:2008qk,Appelquist:2007hu,
Shamir:2008pb,Deuzeman:2008sc,Lucini:2007sa}.

First, the loss of asymptotic freedom is determined by the change of sign in the first coefficient $\beta_0$ of the beta function. This occurs when

\begin{eqnarray} \label{AF}
\sum_{i=1}^{k} \frac{4}{11} T(r_i) N_f(r_i) = C_2(G) \ , \qquad \qquad \text{Loss of AF.}
\end{eqnarray}

Hence for a two color theory with two fundamental flavors the critical number of adjoint Weyl fermions above which one looses asymptotic freedom is $4.50$. Second, we note that at the zero of the beta function we have
\begin{eqnarray}
\sum_{i=1}^{k} \frac{2}{11}T(r_i)N_f(r_i)\left( 2+ \gamma_i \right) = C_2(G) \ .
\end{eqnarray}

 Therefore specifying the value of the anomalous dimensions at the infrared fixed point yields the last constraint needed to construct the conformal window. Having reached the zero of the beta function the theory is conformal in the infrared. For a theory to be conformal the dimension of the non-trivial spinless operators must be larger than one in order to not contain negative norm states \cite{Mack:1975je,Flato:1983te,Dobrev:1985qv}.  Since the dimension of the chiral condensate is $3-\gamma_i$ we see that $\gamma_i = 2$, for all representations $r_i$, yields the maximum possible bound
\begin{eqnarray}\label{Bound}
\sum_{i=1}^{k} \frac{8}{11} T(r_i)N_f(r_i) = C_2(G) \ .
\end{eqnarray}

This implies, for example, that for a two technicolor theory with two fundamental Dirac flavors the critical number of adjoint Weyl fermions needed to reach the bound above on the conformal window  is $1.75$
\footnote{Naively then a two technicolor theory with two fundamental Dirac flavors and one adjoint Weyl fermion would be a good candidate for a NC technicolor theory. However this is hardly the case since the theory equals two flavor supersymmetric QCD but without the scalars. For two colors and two flavors supersymmetric QCD is known to exhibit confinement with chiral symmetry breaking \cite{Intriligator:1995au}. Since the critical number of flavors above which one enters the conformal window is three it will most likely not exhibit NC dynamics. Throwing away the scalars only drives it further away from the NC scenario.} .
The actual size of the conformal window can be smaller than the one determined by the bound above. It may happen, in fact, that chiral symmetry breaking is triggered for a value of the anomalous dimension less than two. If this occurs the conformal window shrinks. Within the ladder approximation \cite{Appelquist:1988yc,{Cohen:1988sq}} one finds that chiral symmetry breaking occurs when the anomalous dimension is close to one. Picking $\gamma_i =1$ we find:
\begin{eqnarray}\label{One}
\sum_{i=1}^{k} \frac{6}{11} T(r_i)N_f(r_i) = C_2(G) \ .
\end{eqnarray}
In this case when considering a two color theory with two fundamental Dirac flavors the critical number of adjoint Weyl flavors is $2.67$. Hence, our candidate for a NC theory with a minimal $S$ parameter has two colors, two fundamental Dirac flavors charged under the electroweak symmetries and two adjoint Weyl fermions. This is the Ultra Minimal NC Technicolor model (UMT).

If it turns out that the anomalous dimension above which chiral symmetry breaking occurs is larger than one we can still use the model just introduced. We will simply break its conformal dynamics by adding masses (anyway needed for phenomenological reasons) for the adjoint fermions.

\section{The Model}
\label{model}
The fermions transforming according to the fundamental representation are arranged into electroweak doublets in the standard way and may be written as:

\begin{eqnarray}
T_L = \left( \begin{array}{c} U \\ D  \end{array}  \right)_L \ , \qquad U_R \ , \ \  D_R
\end{eqnarray}

The additional adjoint Weyl fermions needed to render the theory quasi conformal are denoted as $\lambda^f$ with $f=1,2$. They are not charged under the electroweak symmetries. Also we have suppressed technicolor indices. The theory is anomaly free using the following hypercharge assignment
\begin{eqnarray}
Y(T_L) = 0 \ , \qquad Y(U_R) = \frac{1}{2} \ , \qquad Y(D_R) = -\frac{1}{2} \ , \qquad Y(\lambda^f) = 0 \ ,
\end{eqnarray}
Our notation is such that the electric charge is $Q = T_3 + Y$. Replacing the Higgs sector of the SM with the above technicolor theory the Lagrangian reads:
\begin{eqnarray}
\mathcal{L}_H \rightarrow  -\frac{1}{4}{F}_{\mu\nu}^a {F}^{a\mu\nu} + i\overline{T}_L
\gamma^{\mu}D_{\mu}T_L + i\overline{U}_R \gamma^{\mu}D_{\mu}U_R +
i\overline{D}_R \gamma^{\mu}D_{\mu}D_R + i \overline{\lambda} \overline{\sigma}^{\mu} D_{\mu} \lambda \ ,
\end{eqnarray}
with the technicolor field strength ${F}_{\mu\nu}^a =
\partial_{\mu}{A}_{\nu}^a - \partial_{\nu}{A}_{\mu}^a + g_{TC} \epsilon^{abc} {A}_{\mu}^b
{A}_{\nu}^c,\ a,b,c=1,\ldots,3$. The covariant derivatives for the various fermions are
\begin{eqnarray}
D_{\mu} T_L &=& \left( \partial_{\mu} -i g_{TC} {A}_{\mu}^a \frac{\tau^a}{2} -i g W_{\mu}^a \frac{L^a}{2} \right) T_L \ , \\
D_{\mu}U_R &=& \left( \partial_{\mu} - i g_{TC} A_{\mu}^{a} \frac{\tau^a}{2} - i \frac{g'}{2} B_{\mu} \right)U_R \ , \\
 D_{\mu}D_R &=& \left( \partial_{\mu} - i g_{TC} A_{\mu}^{a} \frac{\tau^a}{2} + i \frac{g'}{2} B_{\mu} \right)D_R \\
 D_{\mu}\lambda^{a,f} &=& \left( \delta^{ac} \partial_{\mu} + g_{TC} A_{\mu}^b \epsilon^{abc} \right) \lambda^{c,f} \ ,
\end{eqnarray}

Here $g_{TC}$ is the technicolor gauge coupling, $g$ is the electroweak gauge coupling and $g'$ is the hypercharge gauge coupling. Also $W_{\mu}^a$ are the electroweak gauge bosons while $B_{\mu}$ is the gauge boson associated to the hypercharge. Both $\tau^a$ and $L^a$ are Pauli matrices and they are the generators of the technicolor and weak gauge groups respectively.

The global symmetries of the theory are most appropriately handled by first arranging the fundamental fermions into a quadruplet of $SU(4)$

\begin{eqnarray}
Q &=& \left( \begin{array}{c}
U_L \\
D_L \\
-i\sigma^2U_R^* \\
-i\sigma^2D_R^*
\end{array} \right) \ .
\end{eqnarray}

Since the fermions belong to pseudo-real and real representations of the gauge group the global symmetry of the theory is enhanced and can be summarized as
\begin{eqnarray}
\begin{array}{c||ccc}\label{symmetries}
 &\qquad SU(4) &\qquad SU(2) &\qquad U(1) \\ \hline\hline
Q &\qquad \fund &\qquad 1 &\qquad -1  \\
\lambda &\qquad 1 &\qquad \fund &\qquad \frac{1}{2}
\end{array}
\end{eqnarray}
The abelian symmetry is anomaly free. Following Ref. \cite{Raby:1979my} the characteristic chiral symmetry breaking scale of the adjoint fermions is larger than that of the fundamental ones since the dimension of the adjoint representation is larger than the dimension of the fundamental representation. We expect, however, the two scales to be very close to each other since the number of fundamental flavors is rather low. In the two-scale technicolor models \cite{Lane:1989ej} the dynamical assumption is instead, that the different scales of the condensates are very much apart from each other.

The global symmetry group $G=SU(4)\times  SU(2) \times U(1)$ breaks to $H= Sp(4) \times SO(2)\times Z_2$. The stability group $H$ is dictated by the (pseudo)reality of the fermion representations and the breaking is triggered by the formation of the following two condensates
\begin{eqnarray}\label{condensate1}
\langle Q_F^{\alpha, c} Q_{F'}^{\beta, c'} \epsilon_{\alpha\beta} \epsilon_{cc'} E_4^{FF'} \rangle &=& -2 \langle \overline{U}_R U_L + \overline{D}_R D_L \rangle \\
\langle \lambda_{f}^{\alpha, k}\lambda_{f'}^{\beta, k'} \epsilon_{\alpha\beta} \delta_{kk'} E_2^{ff'} \rangle &=& -2 \langle \lambda^1 \lambda^2 \rangle \label{condensate2}
\end{eqnarray}
where
\begin{eqnarray}
E_4 = \left( \begin{array}{cc}
0_{2\times2} & \mathbf{1}_{2\times2} \\
-\mathbf{1}_{2\times2} & 0_{2\times2}
\end{array} \right) \ , \qquad E_2 = \left( \begin{array}{cc}
0 & 1 \\
1 & 0
\end{array} \right)
\end{eqnarray}
The flavor indices are denoted with $F,F'=1,\ldots,4$ and $f,f'=1,2$, the spinor indices as $\alpha,\beta=1,2$ and the color indices as $c,c'=1,2$ and $k,k'=1,\ldots,3$. Also the notation is such that $U_L^{\alpha} U_R^{*\beta} \epsilon_{\alpha\beta} = -\overline{U}_RU_L$ and $\lambda^{1,\alpha} \lambda^{2, \beta} \epsilon_{\alpha\beta} = \lambda^1 \lambda^2$. Under the $U(1)$ symmetry $Q$ and $\lambda$ transform as
\begin{equation}
Q \rightarrow e^{-i\alpha}Q \ , \qquad {\rm and } \qquad \lambda \rightarrow e^{-i\frac{\alpha}{2}} \lambda \ ,
\end{equation}
and the two condensates are simultaneously invariant if
\begin{equation}
\alpha = 2 k \pi \ , \qquad {\rm with}~ k ~{\rm an~integer}\ .
\end{equation}
Only the $\lambda$ fields will transform nontrivially under the remaining $Z_2$, i.e.  $\lambda \rightarrow - \lambda$.

\subsection{Low Energy Spectrum}
The relevant degrees of freedom are efficiently collected in two distinct matrices, $M_4$ and $M_2$, which transform as $M_4 \rightarrow g_4M_4g_4^T$ and $M_2 \rightarrow g_2M_2g_2^T$ with $g_4 \in SU(4)$ and $g_2 \in SU(2)$. Both $M_4$ and $M_2$ consist of a composite iso-scalar and its pseudoscalar partner together with the Goldstone bosons and their scalar partners:
\begin{eqnarray}
M_4 &=& \left[ \frac{\sigma_4 + i \Theta_4}{2} + \sqrt{2}\left( i \Pi_4^i+ \tilde{\Pi}_4^i \right) X_4^i \right] E_4 \ , \qquad i=1,\ldots,5 \ , \\
M_2 &=& \left[ \frac{\sigma_2 + i \Theta_2}{\sqrt{2}} + \sqrt{2} \left( i \Pi_2^i+ \tilde{\Pi}_2^i \right) X_2^i \right] E_2 \ , \qquad i=1,2 \ .
\end{eqnarray}

The notation is such that $X_{4}$ and $X_{2}$ are the broken generators of $SU(4)$ and $SU(2)$ respectively. An explicit realization can be found in Appendix \ref{appendixgenerators}. Also $\sigma_{4}$ and $\Theta_{4}$ are the composite Higgs and its pseudoscalar partner while $\Pi_{4}^i$ and $\tilde{\Pi}_{4}^i$ are the Goldstone bosons and their associated scalar partners. {}For SU(2) one simply substitutes the index $4$ with the index $2$.   With the above normalization of the $M$ matrices the kinetic term of each component field is canonically normalized. Under an infinitesimal global symmetry transformation we have:
\begin{eqnarray}
\delta M = i \alpha^a \left( T^a M + M T^{aT} \right) \ .\end{eqnarray}
Here $T$ is the full set of generators of the unbroken group (either SU(4) or SU(2)). With the $\Theta$ and $\tilde{\Pi}^i$ states included the matrices are actually form invariant under $U(4)$ and $U(2)$ with the abelian parts being broken by anomalies. We construct our Lagrangian by considering only the terms preserving the anomaly free U(1) symmetry. As we will see this implies that $\Theta_4$ and $\Theta_2$ are not mass eigensates. In the diagonal basis we will find one massless and one massive state. The massless state corresponds to the $U(1)$ Goldstone boson.

The relation between the composite scalars and the underlying degrees of freedom can be found by first noting that $M_4$ and $M_2$ transform as:
\begin{eqnarray}
M_4^{FF'} \sim Q^FQ^{F'} \ , \qquad M_2^{ff'} \sim \lambda^f \lambda^{f'}
\end{eqnarray}
where both color and spin indices have been contracted. It then follows that the composite states transform as:

\begin{equation}
\begin{array}{rclcrcl}\label{mesons}
\nu_4 +H_4  & \equiv & \sigma_4 \sim \overline{U}U + \overline{D}D & ,~~~~ & \Theta_4 & \sim & i \left( \overline{U}\gamma^5 U + \overline{D} \gamma^5 D \right) \ , \\
\Pi^0 & \equiv & \Pi^3 \sim i \left( \overline{U} \gamma^5 U - \overline{D} \gamma^5 D \right) & ,~~~~ & \tilde{\Pi}^0 & \equiv & \tilde{\Pi}^3  \sim \overline{U}U - \overline{D}D \ , \\
\Pi^+ & \equiv & {\displaystyle \frac{\Pi^1 - i \Pi^2}{\sqrt{2}} \sim i \overline{D} \gamma^5 U} & ,~~~~ & \tilde{\Pi}^+ & \equiv & \frac{\tilde{\Pi}^1 - i \tilde{\Pi}^2}{\sqrt{2}} \sim \overline{D} U\ , \\
\Pi^- & \equiv & \frac{\Pi^1 + i \Pi^2}{\sqrt{2}} \sim i\overline{U} \gamma^5 D & ,~~~~ & \tilde{\Pi}^- & \equiv & \frac{\tilde{\Pi}^1 + i \tilde{\Pi}^2}{\sqrt{2}} \sim \overline{U}D \ , \\
\Pi_{UD} & \equiv & \frac{\Pi^4 + i \Pi^5}{\sqrt{2}} \sim U^T C D & ,~~~~ & \tilde{\Pi}_{UD} & \equiv & \frac{\tilde{\Pi}^4 + i \tilde{\Pi}^5}{\sqrt{2}} \sim iU^T C \gamma^5 D \ , \\
\Pi_{\overline{U}\overline{D}} & \equiv & \frac{\Pi^4 - i \Pi^5}{\sqrt{2}} \sim \overline{U} C \overline{D}^T & ,~~~~ & \tilde{\Pi}_{\overline{U}\overline{D}} & \equiv & \frac{\tilde{\Pi}^4 - i \tilde{\Pi}^5}{\sqrt{2}} \sim i \overline{U} C \gamma^5  \overline{D}^T \ , \\
\end{array}
\end{equation}
and
\begin{equation}
\begin{array}{rclcrcl}\label{mesons}
\nu_2 +H_2  & \equiv & \sigma_2 \sim \overline{\lambda}_D \lambda_D & ,~~~~ & \Theta_2 & \sim & i\overline{\lambda}_D \gamma^5 \lambda_D \ , \\
\Pi_{\lambda \lambda} & \equiv & \frac{\Pi^6-i\Pi^7}{\sqrt{2}} \sim \lambda_D^T C \lambda_D & ,~~~~ & \tilde{\Pi}_{\lambda \lambda} & \equiv & \frac{\tilde{\Pi}^6-i\tilde{\Pi}^7}{\sqrt{2}}  \sim i \lambda_D^T C \gamma_5 \lambda_D\ , \\
\Pi_{\overline{\lambda}\overline{\lambda}} & \equiv & \frac{\Pi^6+i\Pi^7}{\sqrt{2}} \sim \overline{\lambda}_D C \overline{\lambda}_D^T & ,~~~~ & \tilde{\Pi}_{\overline{\lambda}\overline{\lambda}} & \equiv & \frac{\tilde{\Pi}^6+i\tilde{\Pi}^7}{\sqrt{2}}  \sim i \overline{\lambda}_D C \gamma_5 \overline{\lambda}_D^T \ , \\
\end{array}
\end{equation}

Here $U=(U_L,U_R)^T$, $D=(D_L,D_R)^T$ and $\lambda_D=(\lambda^1, - i\sigma^2 {\lambda^2}^{\ast})^T$. Another set of states are the composite fermions
\begin{equation}
\Lambda^f = \lambda^{a,f}  \sigma^{\mu} A_{\mu}^a  \ , \qquad \qquad {f=1,2} \ , \qquad {a=1,2,3} \ .
\end{equation}

To describe the interaction with the weak gauge bosons we embed the electroweak gauge group in $SU(4)$ as done in \cite{Appelquist:1999dq}.  First we note that the following generators
\begin{eqnarray}
L^a = \frac{S^a_4+X^a_4}{\sqrt{2}} = \left( \begin{array}{cc}
\frac{\tau^a}{2} &  \\
 & 0
\end{array} \right) \ , \qquad
R^a = \frac{X^{aT}_4-S^{aT}_4}{\sqrt{2}} = \left( \begin{array}{cc}
0 & \\
 & \frac{\tau^a}{2}
\end{array} \right)
\end{eqnarray}
with $a=1,2,3$ span an $SU(2)_L\times SU(2)_R$ subalgebra. By gauging $SU(2)_L$ and the third generator of $SU(2)_R$ we obtain the electroweak gauge group where the hypercharge is $Y=-R^3$. Then as $SU(4)$ breaks to $Sp(4)$ the electroweak gauge group breaks to the electromagnetic one with the electric charge given by $Q=\sqrt{2}S^3$.
%%%%%%

Due to the choice of the electroweak embedding the weak interactions explicitly reduce the $SU(4)$ symmetry to $SU(2)_L \times U(1)_Y \times U(1)_{TB}$ which is further broken to $U(1)_{\rm em} \times U(1)_{TB}$ via the technicolor interactions. $U(1)_{TB}$ is the technibaryon number and its generator corresponds to the $S_4^4$ diagonal generator (see appendix \ref{appendixgenerators}). The remaining $SU(2)\times U(1)$ spontaneously break, only via the (techni)fermion condensates, to $SO(2)\times Z_2$. We prefer to indicate $SO(2)$ with $U(1)_{T\lambda}$. We summarize some of the relevant low-energy technihadronic states according to the final unbroken symmetries in Table \ref{symmetries}.
\begin{table}
\begin{eqnarray}
\begin{array}{c||ccccc}
 &\qquad SU(2)_L &\qquad U(1)_{\rm em} &\qquad U(1)_{ TB} &\qquad U(1)_{T\lambda}& \qquad  Z_2 \\ \hline\hline
H_4,\ \Theta_4 & \qquad  1 &\qquad 0  &\qquad 0 & \qquad 0 & \qquad 0 \\
\stackrel{\rightarrow}{\Pi},\  \stackrel{\rightarrow}{\tilde{\Pi}} &\qquad 3 &\qquad +1,0,-1 &\qquad 0 & \qquad 0 & \qquad 0  \\
\Pi_{UD},\  \tilde{\Pi}_{UD}&\qquad  1 &\qquad 0  &\qquad \frac{1}{\sqrt{2}} & \qquad 0 & \qquad 0 \\
\Pi_{\overline{U}\overline{D}} ,\  \tilde{\Pi}_{\overline{U}\overline{D}} & \qquad  1 &\qquad 0  &\qquad -\frac{1}{\sqrt{2}} & \qquad 0 & \qquad 0 \\
H_2,\ \Theta_2 & \qquad  1 &\qquad 0  &\qquad 0 & \qquad 0 & \qquad 0 \\
\Pi_{\lambda\lambda},\ \tilde{\Pi}_{\lambda\lambda} & \qquad  1 &\qquad 0  &\qquad 0 & \qquad 1 & \qquad 0 \\
\Pi_{\overline{\lambda}\overline{\lambda}},\ \tilde{\Pi}_{\overline{\lambda}\overline{\lambda}} & \qquad  1 &\qquad 0  &\qquad 0 & \qquad -1 & \qquad 0 \\
\Lambda_D&\qquad  1 &\qquad 0  &\qquad 0 & \qquad \frac{1}{2} & \qquad -1
\end{array} \nonumber
\end{eqnarray}
\caption{Summary table of the relevant low-energy technihadronic states for UMT. We display their $SU(2)_L$ weak interaction charges together with their electromagnetic ones. We also show the remaining global symmetries.}
\label{symmetries}
\end{table}
We have arranged the composite fermions into a Dirac fermion
\begin{equation}
\Lambda_D=
\left(
\begin{array}{c}
  \Lambda^1     \\
  -i\sigma^2 {\Lambda^2}^{\ast}
\end{array}
\right) \ .
\end{equation}

Except for the triplet of Goldstone bosons charged under the electroweak symmetry the rest of the states are electroweak neutral.  In the unitary gauge the $\vec{\Pi}$ states become the longitudinal components of the massive electroweak gauge bosons. $\Pi_{UD}$ ($\tilde{\Pi}_{UD}$) is a pseudoscalar(scalar) diquark charged under the technibaryon number $U(1)_{TB}$ while $\Pi_{\lambda\lambda}$ (${\tilde\Pi}_{\lambda\lambda}$) is charged under the $U(1)_{T\lambda}$. $\Lambda_D$ is the composite fermionic state charged under
both $U(1)_{T\lambda}$ and $Z_2$.

The technibaryon number $U(1)_{TB}$  is anomalous due to the presence of the weak interactions:
\begin{equation}
\partial_{\mu} J^{\mu}_{TB} = \frac{1}{2\sqrt{2}} \frac{g^2}{32 \pi^2} \epsilon_{\mu\nu\rho\sigma} W^{\mu \nu} {W}^{\rho\sigma}  \ , \quad {\rm and} \quad
%%% (Explanation)
%%% 1/(2\sqrt{2}) (From U(1)_TB charge) x 1/2 (Tr[TT]) x 2(Number of technicolors) x \frac{g^2}{32 \pi^2} \epsilon_{\mu\nu\rho\sigma} W^{\mu \nu} {W}^{\rho\sigma}
%%%%
J^{\mu}_{TB} =  \frac{1}{2\sqrt{2}}\left( \bar{U} \gamma^{\mu}U +
 \bar{D} \gamma^{\mu}D \right) \ .
\end{equation}

\subsection{Linear Lagrangian}

With the above discussion of the electroweak embedding the covariant derivative for $M_4$ is:
\begin{eqnarray}
D_{\mu} M_4 &=& \partial_{\mu} M_4 - i \left[ G_{\mu} M_4 + M_4 G_{\mu}^T  \right] \ , \qquad G_{\mu} =
\left( \begin{array}{cc}
g W_{\mu}^a \frac{\tau^a}{2} & 0 \\
0 & -g' B_{\mu} \frac{\tau^3}{2}
\end{array} \right)  \ .
\end{eqnarray}

We are now in a position to write down the effective Lagrangian. It contains the kinetic terms and a potential term:
\begin{eqnarray}
\mathcal{L} &=& \frac{1}{2} \text{Tr} \left[ D_{\mu} M_4 D^{\mu} M_4^{\dagger} \right] + \frac{1}{2} \text{Tr} \left[ \partial_{\mu} M_2 \partial^{\mu} M_2^{\dagger} \right] - \mathcal{V} \left( M_4, M_2 \right)
\end{eqnarray}
where the potential is:
\begin{eqnarray}
\mathcal{V} \left( M_4, M_2 \right) &=& -\frac{m_4^2}{2} \text{Tr}\left[ M_4M_4^{\dagger} \right] + \frac{\lambda_4}{4}\text{Tr}\left[ M_4M_4^{\dagger} \right]^2 + \lambda_4' \text{Tr} \left[ M_4M_4^{\dagger}M_4M_4^{\dagger} \right] \\
&&
-\frac{m_2^2}{2} \text{Tr}\left[ M_2M_2^{\dagger} \right] + \frac{\lambda_2}{4}\text{Tr}\left[ M_2M_2^{\dagger} \right]^2 + \lambda_2' \text{Tr} \left[ M_2M_2^{\dagger}M_2M_2^{\dagger} \right] \\
&&
+ \frac{\delta}{2}\text{Tr}\left[ M_4M_4^{\dagger} \right] \text{Tr}\left[ M_2M_2^{\dagger} \right] +4\delta' \left[ \left( \det M_2 \right)^2 \text{Pf}\ M_4 + \text{h.c.} \right] \ .
\end{eqnarray}

Once $M_4$ develops a vacuum expectation value the electroweak symmetry breaks and three of the eight Goldstone bosons - $\Pi^0,\ \Pi^+$ and $\Pi^-$ - will be eaten by the massive gauge bosons. In terms of the parameters of the theory the vacuum states $\langle \sigma_4 \rangle = v_4$ and $\langle \sigma_2 \rangle = v_2$ which minimize the potential are a solution of the two coupled equations
\begin{eqnarray}\label{vacua}
0 &=& -m_4^2 - \left(\delta + \delta' v_2^2 \right) v_2^2 + \left( \lambda_4 + \lambda_4' \right) v_4^2 \ , \\
0 &=& -m_2^2 - \left( \delta + 2\delta' v_2^2 \right) v_4^2 + \left( \lambda_2 + 2\lambda_2' \right) v_2^2 \ .
\end{eqnarray}

Expanding around the symmetry breaking vacua all of the Goldstone bosons scalar partners are seen to be mass eigenstates with masses
\begin{eqnarray}
M^2_{\tilde{\Pi}^0} = M^2_{\tilde{\Pi}^{\pm}} = M^2_{\tilde{\Pi}_{UD}} = 2 \left( \lambda_4' v_4^2 + \delta' v_2^4 \right) \ , \qquad M^2_{\tilde{\Pi}_{\lambda\lambda}} = 4v_2^2 \left( \lambda_2'+ \delta' v_4^2 \right) \ ,
\end{eqnarray}
while the Goldstone bosons which are not eaten by the massive gauge bosons of course have vanishing mass $M^2_{\Pi_{UD}} = M^2_{\Pi_{\lambda\lambda}} = 0$. Here the vacuum expectation values $v_4$ and $v_2$ are solutions to Eq.~(\ref{vacua}). Due to the presence of the determinant/Pfaffian term in the potential the remaining states are not mass eigenstates. Specifically $H_4$ and $H_2$ and their associated pseudoscalar partners will mix. In the diagonal basis we find the following mass eigenstates:
\begin{equation}
\begin{array}{rclcrcl}\label{mesons}
\Theta  & \equiv &  \sin (\alpha)\ \Theta_4 + \cos (\alpha)\ \Theta_2 & ,~~~~ & M_{\Theta}^2 & = & 0 \ , \\
\tilde{\Theta}  & \equiv & \cos (\alpha)\ \Theta_4 - \sin (\alpha)\ \Theta_2  & ,~~~~ & M_{\tilde{\Theta}}^2 & = & 2 \delta' v_2^2 \left( v_2^2 + 4 v_4^2 \right) \ , \\
H_-  & \equiv & \sin (\beta)\ H_4 + \cos (\beta)\ H_2 & ,~~~~ & M^2_{H_-} & = &  m_2^2 +m_4^2 + k_-  \ , \\
H_+  & \equiv & \cos (\beta)\ H_4 -  \sin(\beta) \ H_2 & ,~~~~ & M^2_{H_+} & = & m_2^2 +m_4^2 + k_+ \ , \\
\end{array}
\end{equation}
with
\begin{eqnarray}
\tan (2\alpha) = \frac{4v_4v_2}{v_2^2-4v_4^2} \ , \qquad \tan (2\beta) = \frac{2v_2v_4\left( \delta + 2 \delta' v_2^2  \right)}{m_2^2 - m_4^2 + \delta v_4^2 - \left( \delta + \delta' v_2^2 \right) v_2^2}  \ ,
\end{eqnarray}
\begin{eqnarray}
k_{\pm}= \left( \delta + \delta'v_2^2 \right) v_2^2 + \delta v_4^2 \pm \left[ \left( m_4^2 - m_2^2 + \left( \delta + \delta' v_2^2 \right)v_2^2 - \delta v_4^2 \right)^2 + \left( 2v_2v_4 \left( \delta + \delta' v_2^2 \right) \right)^2 \right]^{\frac{1}{2}}
\end{eqnarray}
Note that we have one massless state $\Theta$ which we identify with the original $U(1)$ Goldstone boson while $\tilde{\Theta}$ is massive. In the limit $\delta' \rightarrow 0$ both states are massless and at the classical level the global symmetry is enhanced to $U(4)\times U(2)$.

For the model to be phenomenologically viable some of the Goldstones must acquire a mass. This is typically addressed by extending the technicolor interactions (ETC). A review of the major models is given by Hill and Simmons
\cite{Hill:2002ap}. At the moment there is not yet a consensus on
which ETC is the best. Here we parameterize the ETC interactions by adding at the effective Lagrangian level the operators needed to give the dangerous Goldstone bosons an explicit mass term.

The effective ETC Lagrangian breaks the global $SU(4)\times SU(2) \times U(1)$ symmetry. The $SU(4)$ generator commuting with the $SU(2)_L \times SU(2)_R$ generators is $B_4 = 2\sqrt{2} S_4^4$. To construct, at the effective Lagrangian level, the interesting ETC terms we find it useful to split
$M_4$ ($M_2$) -- form invariant under $U(4)$ ($U(2)$) --  as follows:
\begin{equation}
M_4 = \tilde{M}_4  + i P_4 \ , \qquad  {\rm and} \qquad M_2 = \tilde{M}_2  + i P_2 \ ,
\end{equation}
with
\begin{eqnarray}
\tilde{M}_4 &=& \left[ \frac{\sigma_4}{2} + i\sqrt{2}  \Pi_4^i   X_4^i \right]   E_4 \ , \quad P_4 = \left[ \frac{ \Theta_4}{2}  -i\sqrt{2} \tilde{\Pi}_4^i X_4^i \right] E_4 \ , \qquad i=1,\ldots,5 \ , \\
\tilde{M}_2 &=& \left[ \frac{\sigma_2 }{\sqrt{2}} + i\sqrt{2}  \Pi_2^i  X_2^i \right] E_2 \ , \quad P_2 = \left[ \frac{ \Theta_2}{\sqrt{2}}  -i\sqrt{2} \tilde{\Pi}_2^i  X_2^i \right] E_2 \ , \qquad i=1,2 \ .\end{eqnarray}
$\tilde{M}_4 $ ($\tilde{M}_2$) as well as $P_4$ ($P_2$) are separately $SU(4)$ ($SU(2)$) form invariant. A set of operators able to give masses to the electroweak neutral Goldstone bosons is:
\begin{eqnarray}
\mathcal{L}_{ETC} &=&  \frac{m_{4,ETC}^2}{4} \text{Tr} \left[ \tilde{M}_4B_4 \tilde{M}_4^{\dagger} B_4 + \tilde{M}_4 \tilde{M}_4^{\dagger} \right] + \frac{m_{2,ETC}^2}{4} \text{Tr} \left[ \tilde{M}_2B_2 \tilde{M}_2^{\dagger} B_2 + \tilde{M}_2 \tilde{M}_2^{\dagger} \right] \nonumber \\
&& - m_{1,ETC}^2 \left[ \text{Pf}\ P_4 + \text{Pf}\ P_4^{\dagger} \right] - \frac{m_{1,ETC}^2}{2} \left[ \det(P_2) + \det(P_2^{\dagger}) \right] \ ,
\end{eqnarray}
where $B_2=2S_2^1$. The spectrum is:
\begin{eqnarray}
M^2_{\Pi_{UD}} = m_{4,ETC}^2 \ , \qquad M^2_{\Pi_{\lambda\lambda}} = m_{2,ETC}^2 \ , \qquad M^2_{\Theta} = m_{1,ETC}^2 \ ,
\end{eqnarray}
for the Goldstone bosons that are not eaten by the massive vector bosons and:
\begin{eqnarray}
M^2_{\tilde{\Pi}_{UD}} &=& M^2_{\tilde{\Pi}^0} = M^2_{\tilde{\Pi}^{\pm}} = 2\left( \lambda'_4 v_4^2 + \delta' v_2^4 \right) + m^2_{1,ETC} \ , \\
M^2_{\tilde{\Pi}_{\lambda\lambda}} &=& 4v_2^2 \left( \lambda_2' + \delta' v_4^2 \right) + m^2_{1,ETC} \ , \\
M_{\tilde{\Theta}}^2 &=& 2\delta' v_2^2 \left( v_2^2 + 4v_4^2 \right) + m_{1,ETC}^2 \ ,
\end{eqnarray}
for the pseudoscalar and scalar partners. The masses of the two Higgs particles $H_+$ and $H_-$ are unaffected by the addition of the ETC low energy operators.

\subsection{Non-Linear Lagrangian}

In constructing the non-linear effective theory of the associated Goldstone bosons we shall consider the elements of the global symmetry $G$ as $6\times 6$ matrices.
The generators of $SU(4)$ sit in the upper left corner while the generators of $SU(2)$ sit in the lower right corner. The generator of $U(1)$ is diagonal. We divide the nineteen generators of $G$ into the eleven that leave the vacuum invariant $S$ and the eight that do not $X$. An explicit realization of $S$ and $X$ can be found in Appendix \ref{appendixgenerators}.

An element of the coset space $G/H$ is parameterized by
\begin{eqnarray}
\mathcal{V}(\xi) = \exp \left( i \xi^iX^i \right) E \ ,
\end{eqnarray}
where
\begin{eqnarray}
E= \left( \begin{array}{cc}
E_4 & \\
  & E_2
\end{array}\right) \ , \qquad  \xi^i X^i = \sum_{i=1}^{5} \frac{\Pi^iX^i}{F_{\pi}} + \sum_{i=6}^{7} \frac{\Pi^iX^i}{\tilde{F}_{\pi}} + \frac{\Pi^8 X^8}{\hat{F}_{\pi}} \ .
\end{eqnarray}
The Goldstone bosons are denoted as $\Pi^i,\ i=1,\ldots,8$ and $F_{\pi}, \tilde{F}_{\pi}$ and $\hat{F}_{\pi}$ are the related Goldstone boson decay constants. Since the entire global symmetry $G$ is expected to break approximately at the same scale we also expect the three decay constants to have close values. The element $\mathcal{V}$ of the coset space transforms non-linearly
\begin{eqnarray} \label{trans}
\mathcal{V}(\xi) \rightarrow g \mathcal{V}(\xi) h^{\dagger}(\xi, g)
\end{eqnarray}
where $g$ is an element of $G$ and $h$ is an element of $H$. To describe the Goldstone bosons interaction with the weak gauge bosons we embed the electroweak gauge group in $SU(4)$ as done above and also in \cite{Appelquist:1999dq}.
With the embedding of the electroweak gauge group in hand it is appropriate to introduce the hermitian, algebra valued, Maurer-Cartan one-form
\begin{eqnarray}
\omega_{\mu} = i \mathcal{V}^{\dagger} D_{\mu} \mathcal{V}
\end{eqnarray}
where the electroweak covariant derivative is
\begin{eqnarray}
D_{\mu} \mathcal{V} = \partial_{\mu} \mathcal{V} -iG_{\mu} \mathcal{V} \ , \qquad G_{\mu} = \left( \begin{array}{ccc}
gW_{\mu}^{a} \frac{\tau^a}{2} & & \\
 & -g'B_{\mu}\frac{\tau^3}{2}  &  \\
 & & 0
\end{array} \right) \ .
\end{eqnarray}
From the above transformation properties of $\mathcal{V}$ it is clear that $\omega_{\mu}$ transforms as
\begin{eqnarray}
\omega_{\mu} \rightarrow h(\xi, g) \omega_{\mu} h^{\dagger}(\xi, g) + h(\xi, g) \partial_{\mu} h^{\dagger}(\xi, g) \ .
\end{eqnarray}
With $\omega_{\mu}$ taking values in the algebra of $G$ we can decompose it into a part $\omega_{\mu}^{\parallel}$ parallel to $H$ and a part $\omega_{\mu}^{\perp}$ orthogonal to $H$
\begin{eqnarray}
\omega_{\mu}^{\parallel} = 2S^a \text{Tr}\left[ S^a \omega_{\mu} \right] \ , \qquad \omega_{\mu}^{\perp} = 2X^i \text{Tr} \left[ X^i \omega_{\mu} \right] \ .
\end{eqnarray}
It is clear that $\omega_{\mu}^{\parallel}$ ($\omega_{\mu}^{\perp}$) is an element of the algebra of $H$ ($G/H$) since it is a linear combination of $S^a$ ($X^i$). They have the following transformation properties
\begin{eqnarray}
\omega_{\mu}^{\parallel} \rightarrow h(\xi, g) \omega_{\mu}^{\parallel} h^{\dagger}(\xi, g) + h(\xi, g) \partial_{\mu} h^{\dagger}(\xi, g) \ , \qquad \omega_{\mu}^{\perp} \rightarrow h(\xi, g) \omega_{\mu}^{\perp} h^{\dagger}(\xi, g)
\end{eqnarray}

We are now in a position to construct the non-linear Lagrangian. We shall only consider terms containing at most two derivatives. By noting that the generator $X^8$ corresponding to the broken $U(1)$ is not traceless we can also write a double-trace term besides the standard one-trace term:
\begin{eqnarray}
\mathcal{L} = \text{Tr} \left[a \omega_{\mu}^{\perp} \omega^{\mu\perp}  \right] + b \text{Tr} \left[ \omega_{\mu}^{\perp} \right] \text{Tr} \left[ \omega^{\mu \perp} \right] \ ,
\end{eqnarray}
The coefficients $a=\text{diag}\left(F_{\pi}^2, F_{\pi}^2, F_{\pi}^2, F_{\pi}^2, \tilde{F}_{\pi}^2, \tilde{F}_{\pi}^2 \right)$ and $b=\frac{\hat{F}_{\pi}^2}{2} - \frac{4F_{\pi}^2}{9} - \frac{\tilde{F}_{\pi}^2}{18}$ are chosen such that the kinetic term is canonically normalized:
\begin{eqnarray}
\mathcal{L} = \frac{1}{2} \sum_{i=1}^8 \partial_{\mu} \Pi^i \partial^{\mu} \Pi^i + \ldots \ .
\end{eqnarray}
We conclude this section by connecting the linear and non-linear theories
\begin{eqnarray}
F_{\pi}^2 = \frac{v_4^2}{2} \ , \qquad \tilde{F}_{\pi}^{2} = v_2^2 \ , \qquad \hat{F}_{\pi}^2 = \frac{1}{9} \left( 4v_4^2 + v_2^2 \right) \ .
\end{eqnarray}

\section{The TIMP}
\label{timp}

Technicolor models are capable of
providing interesting DM candidates. This is so since the  new
strong
interactions confine techniquarks in technimeson and technibaryon
bound states. The spin of the technibaryons depends on the
representation according to which the technifermions transform, as well as
the number of flavors and colors. The lightest technimeson is
short-lived, thus evading BBN constraints \cite{Chivukula:1989qb}, while the lightest
technibaryon can be stable and may posses a dynamical mass of the order
\begin{equation}
 m_{TB}  \sim  1-2\ {\rm TeV} \ .
\end{equation}
If the lighest technibaryon is only weakly interacting and electrically neutral it can be a DM candidate as first suggest by Nussinov \cite{Nussinov:1985xr}. This proposal has been further analyzed in
\cite{Barr:1990ca,Gudnason:2006yj}. One of the interesting properties of this kind of DM candidate is that it is
possible to understand the observed ratio of the dark to
luminous mass of the universe. This occurs when the technibaryon relic density is caused by
 a technibaryon number (TB) asymmetry \cite{Nussinov:1985xr,Barr:1990ca,Gudnason:2006yj} like for the ordinary baryon (B). If the latter is
due to a net Baryon - Lepton ($B-L$) asymmetry generated at some high energy scale, this
would subsequently be distributed among {all} electroweak
doublets via SM fermion-number violating processes at
temperatures above the electroweak scale
\cite{Shaposhnikov:1991cu,Kuzmin:1991ft,Shaposhnikov:1991wi},  thus
 generating a technibaryon asymmetry as well. To avoid
experimental constraints the technibaryon should be a complete singlet under the electroweak
interactions \cite{Barr:1990ca,Dietrich:2006cm}. These kinds of particles are Technicolor Interacting Massive Particles (TIMP)s which are hard to detect \cite{Bagnasco:1993st,Gudnason:2006yj,Kouvaris:2008hc} in current earth-based
experiments  such as CDMS
\cite{Ahmed:2008eu}.
Other possibilities have been envisioned in
\cite{Kouvaris:2007iq,Khlopov:2007ic} and astrophysical
effects investigated in \cite{Kouvaris:2007ay}. One can alternatively
obtain DM from possible technicolor-related new sectors \cite{Kainulainen:2006wq}.   In \cite{Sannino:2008ha} the reader will
find an up-to-date summary of the recent efforts in this direction.

Our extension of the SM naturally provides a {\it novel} type of TIMP, i.e. a  di-techniquark,
with the following unique features:
\begin{itemize}
\item It is a quasi-Goldstone of the underlying gauge
theory receiving a mass term only from interactions not present in the
technicolor theory per se.
\item The lightest technibaryon is a {\it singlet}
with respect to weak interactions.
\item Its relic density can be related to
the SM lepton number over the baryon number if the asymmetry is produced above the
eletroweak phase transition.
\end{itemize}

In appendix  \ref{darkmatter} we provide a much detailed model
computation of the ratio $TB/B$ making use of
the chemical equilibrium conditions and the sphaleron processes active
around the electroweak phase transition.

In the approximation where also the top quark is considered massless around
the electroweak phase transition (we have also checked that the effects of the top mass do not change our results) the $TB/B$ is independent of
the order of the electroweak phase transition and reads
\begin{eqnarray}
- \frac{\sqrt{2}\cdot TB}{B} &=& \frac{\sigma}{2} \left( 3 + \xi \right) \ ,
\end{eqnarray}
where  $\sigma\equiv \sigma_U = \sigma_D$ is the statistical function for the  techniquarks.  The $U$ and $D$ constitutent-type masses are assumed to be dynamically generated and equal.  $\xi = L/B$ is the SM lepton over the baryon number.

If DM is identified with the lightest technibaryon in our model the
ratio of the dark to baryon mass of the universe is
\begin{equation}
\frac{\Omega_{TB}}{\Omega_B} = \frac{m_{TB}}{m_p}  \frac{\widetilde{TB}}{B} \ ,
\end{equation}
with $m_{TB}$ the technibaryon mass and $\widetilde{TB}=- {\sqrt{2}
TB}$ the technibaryon number normalized in such a way that it is minus one for the lightest state.

 The bulk of the mass of the lightest technibaryon is not due to the technicolor interactions as it was in the original proposal  \cite{Nussinov:1985xr,Barr:1990ca}. This is similar to the
case studied in \cite{Gudnason:2006yj}. The interactions providing mass to the techibaryon are the SM interactions per se and ETC. The main effect of these interactions
will be in the strength and the order of the electroweak phase transition as shown in
\cite{Cline:2008hr}.

\begin{figure}[b]
{\includegraphics[height=6.5cm,width=7.82cm]{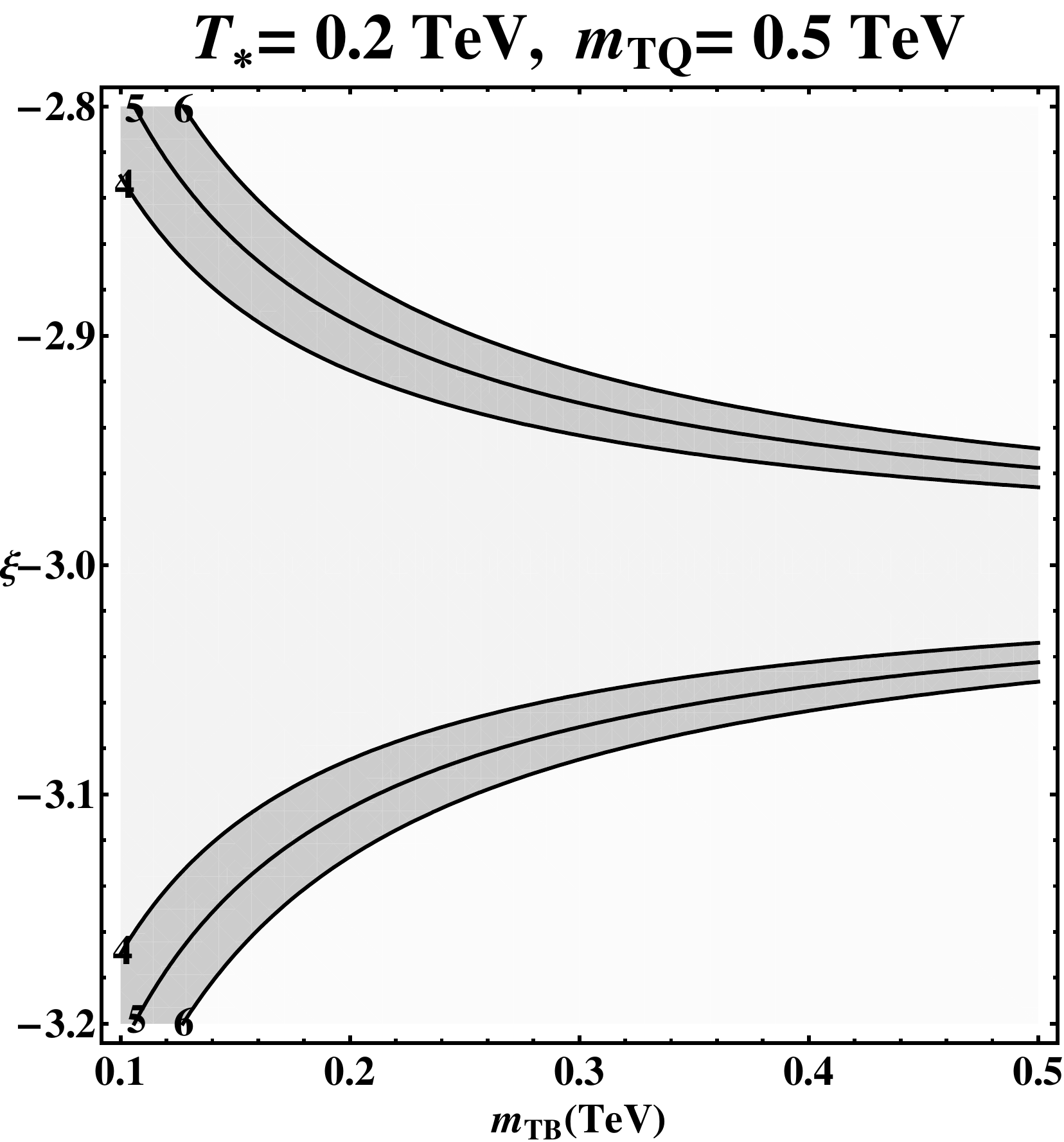}\hspace{0.8cm}\includegraphics[height=6.5cm,width=7.82cm]{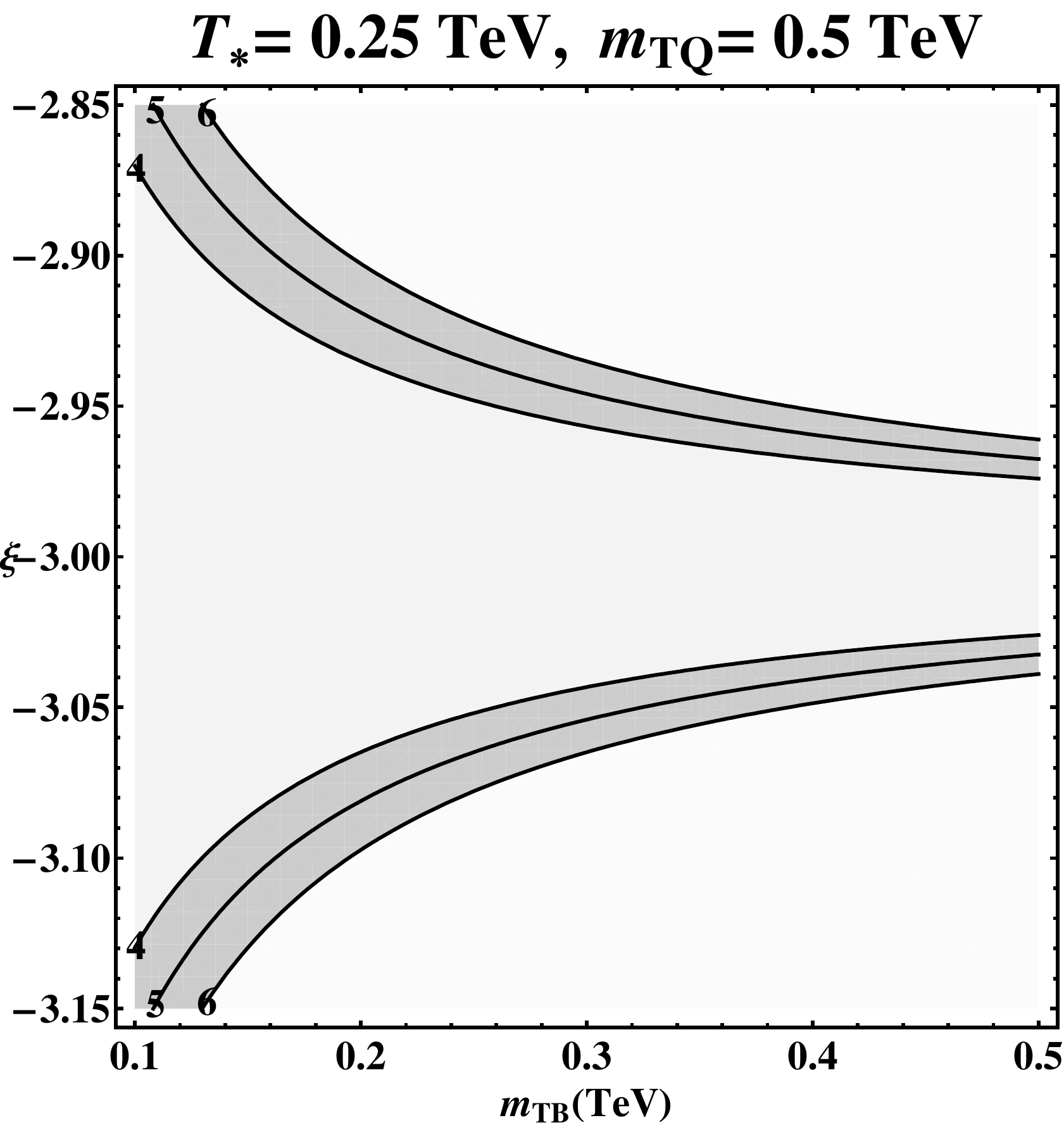}\hspace{0.8cm}\includegraphics[height=6.5cm,width=7.82cm]{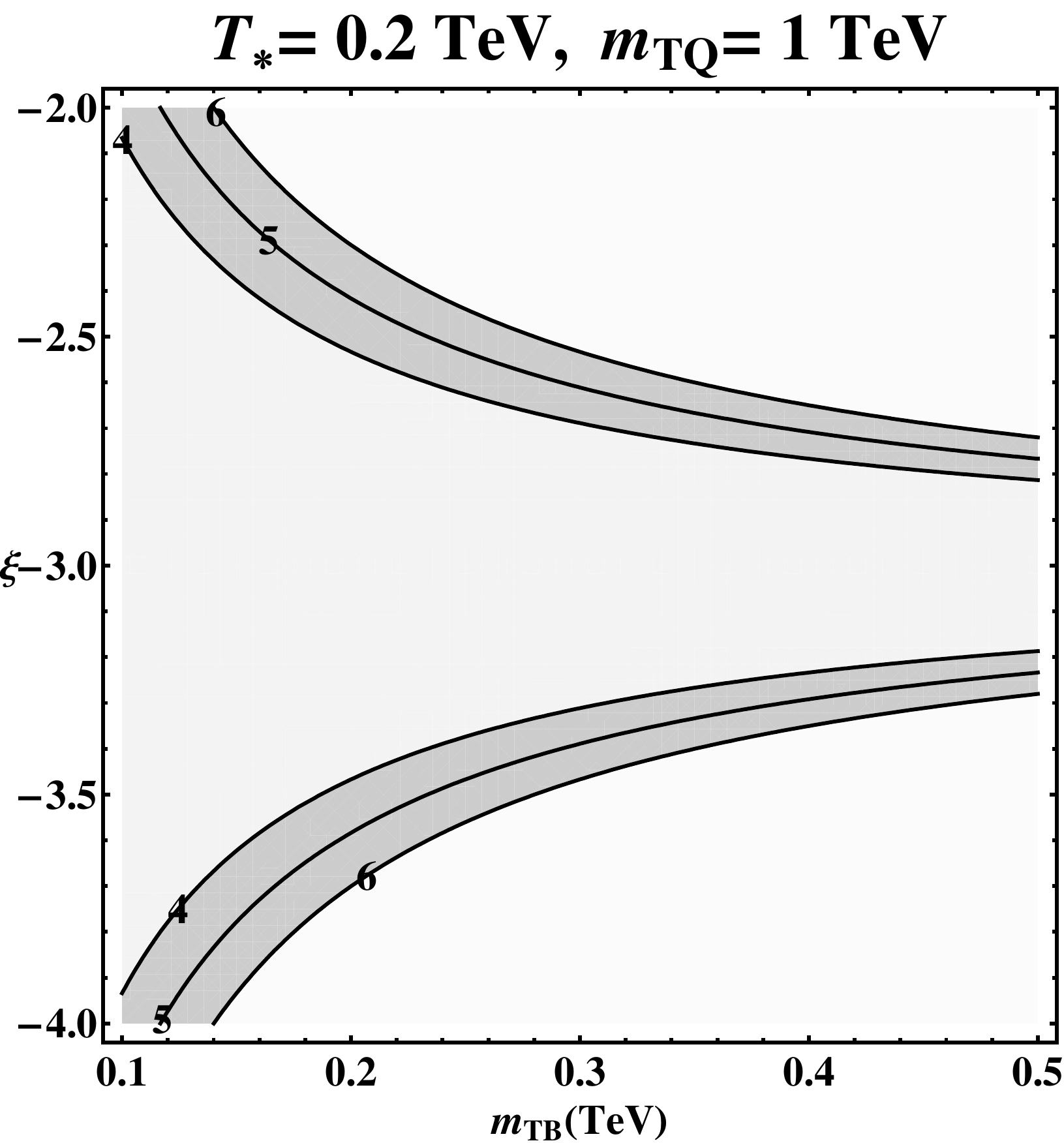}\hspace{0.8cm}\includegraphics[height=6.5cm,width=7.82cm]{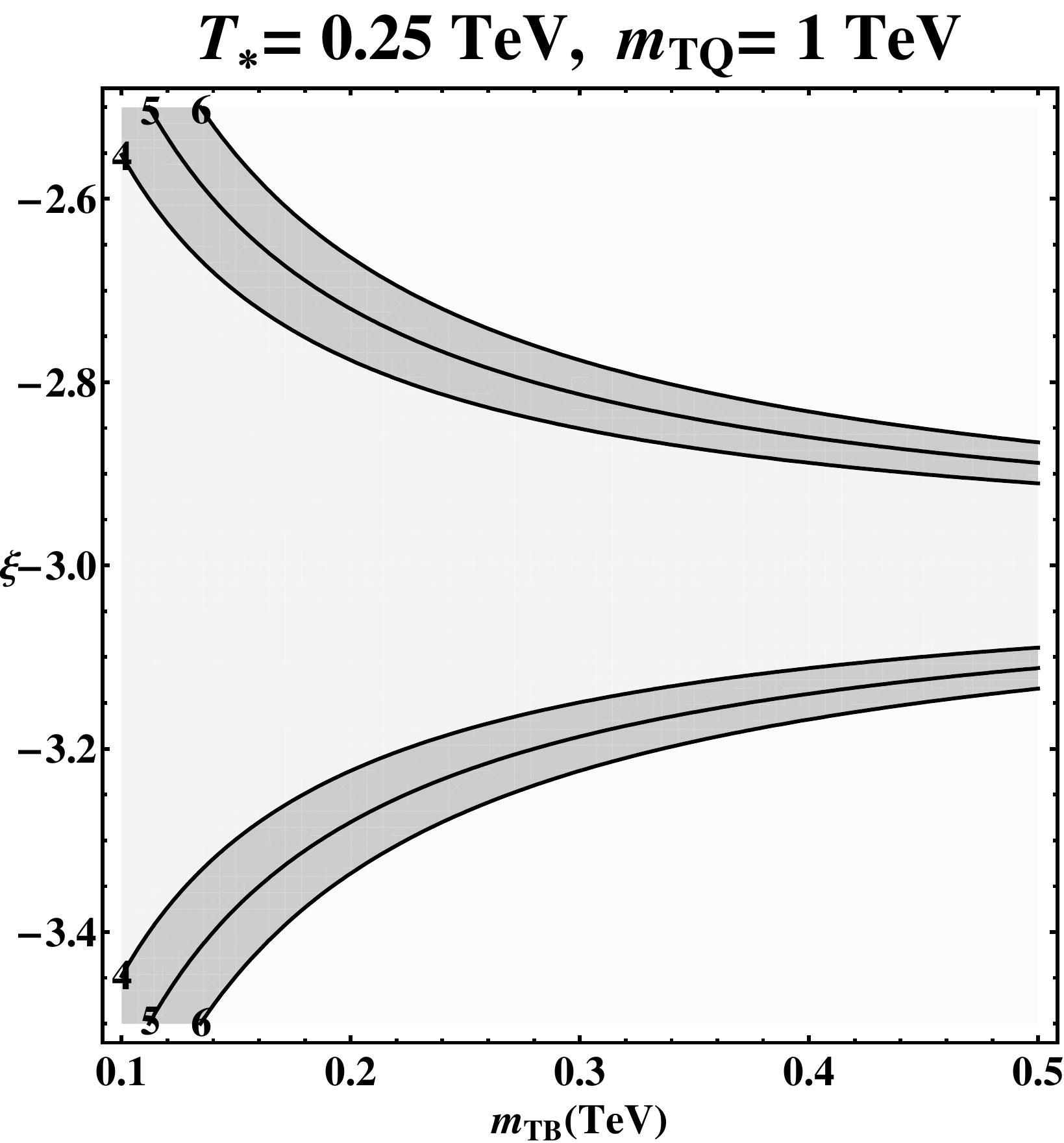}\hspace{0.8cm}}
\end{figure}

\begin{figure}[t]
{\includegraphics[height=6.5cm,width=7.82cm]{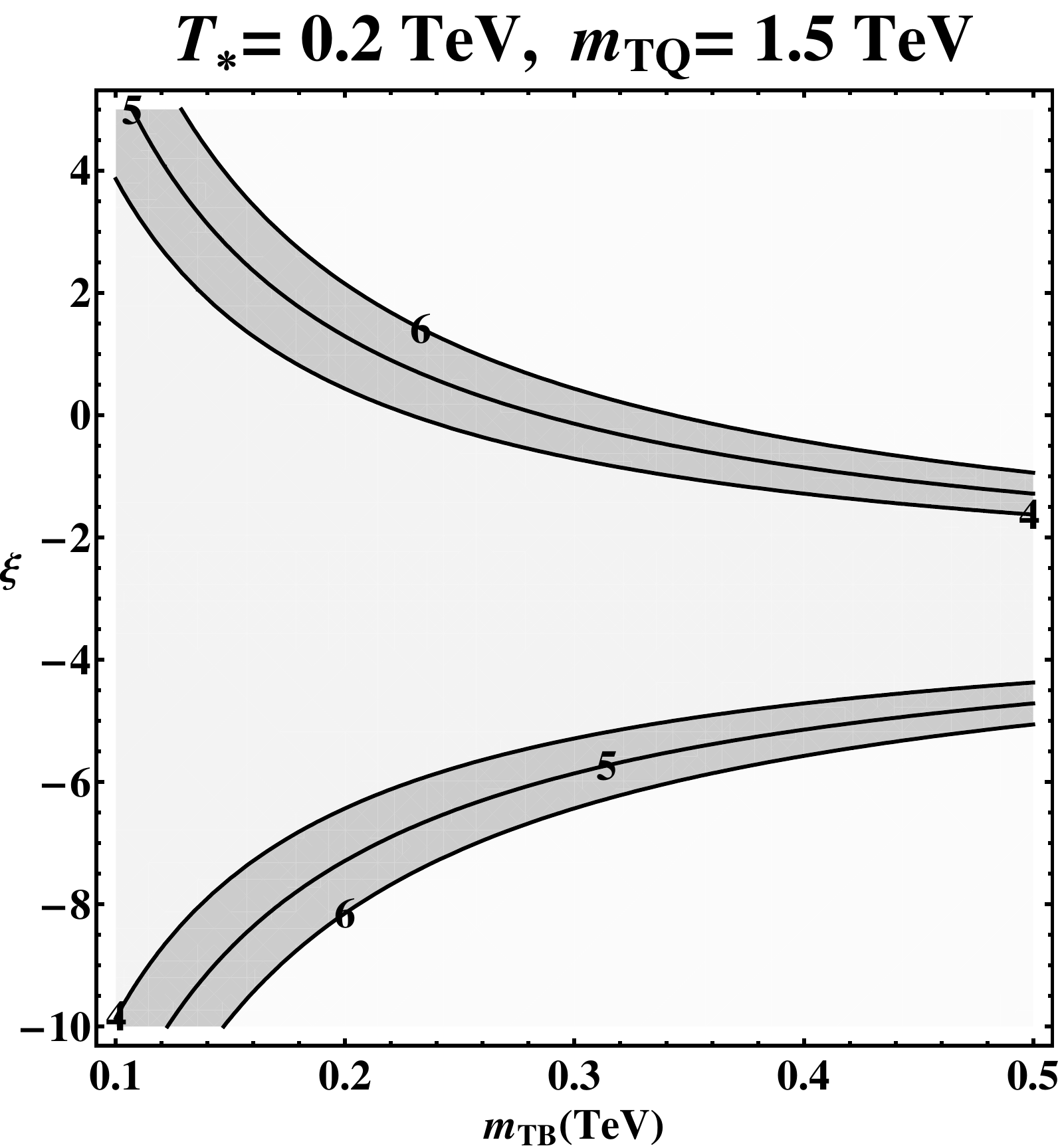}\hspace{0.8cm}\includegraphics[height=6.5cm,width=7.82cm]{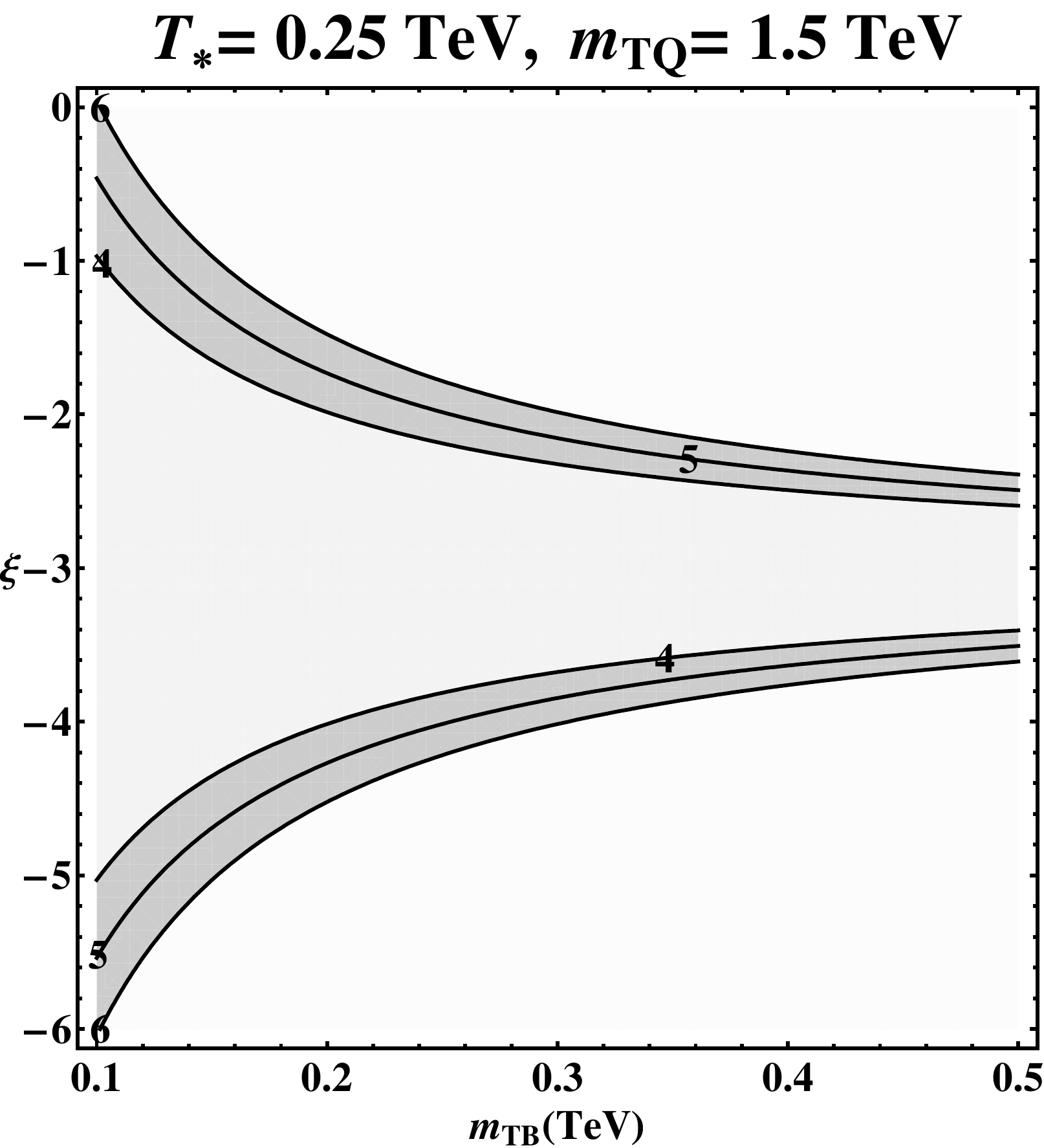}\vspace{.8cm}
\includegraphics[height=6.5cm,width=7.82cm]{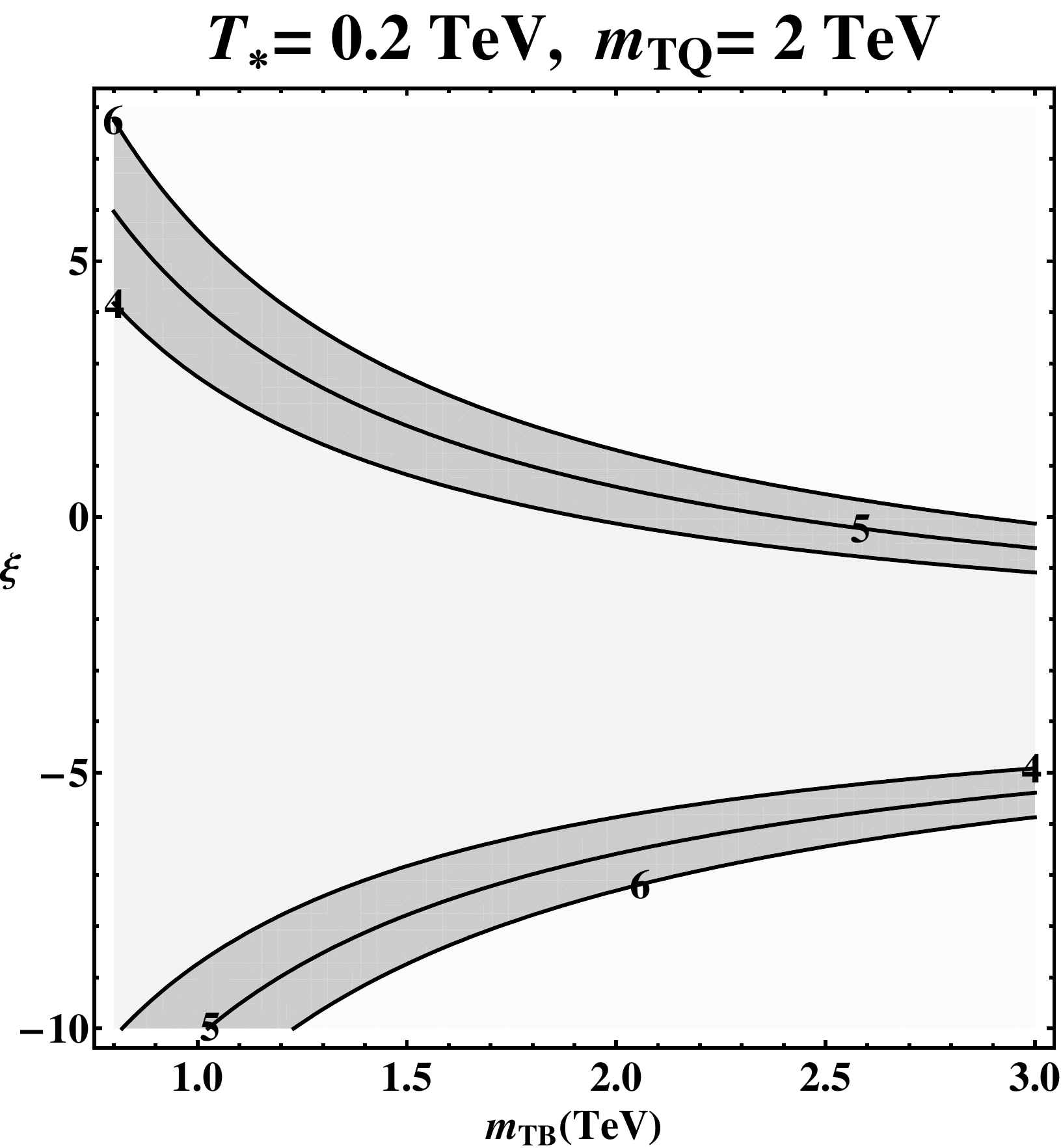}\hspace{0.8cm}\includegraphics[height=6.5cm,width=7.82cm]{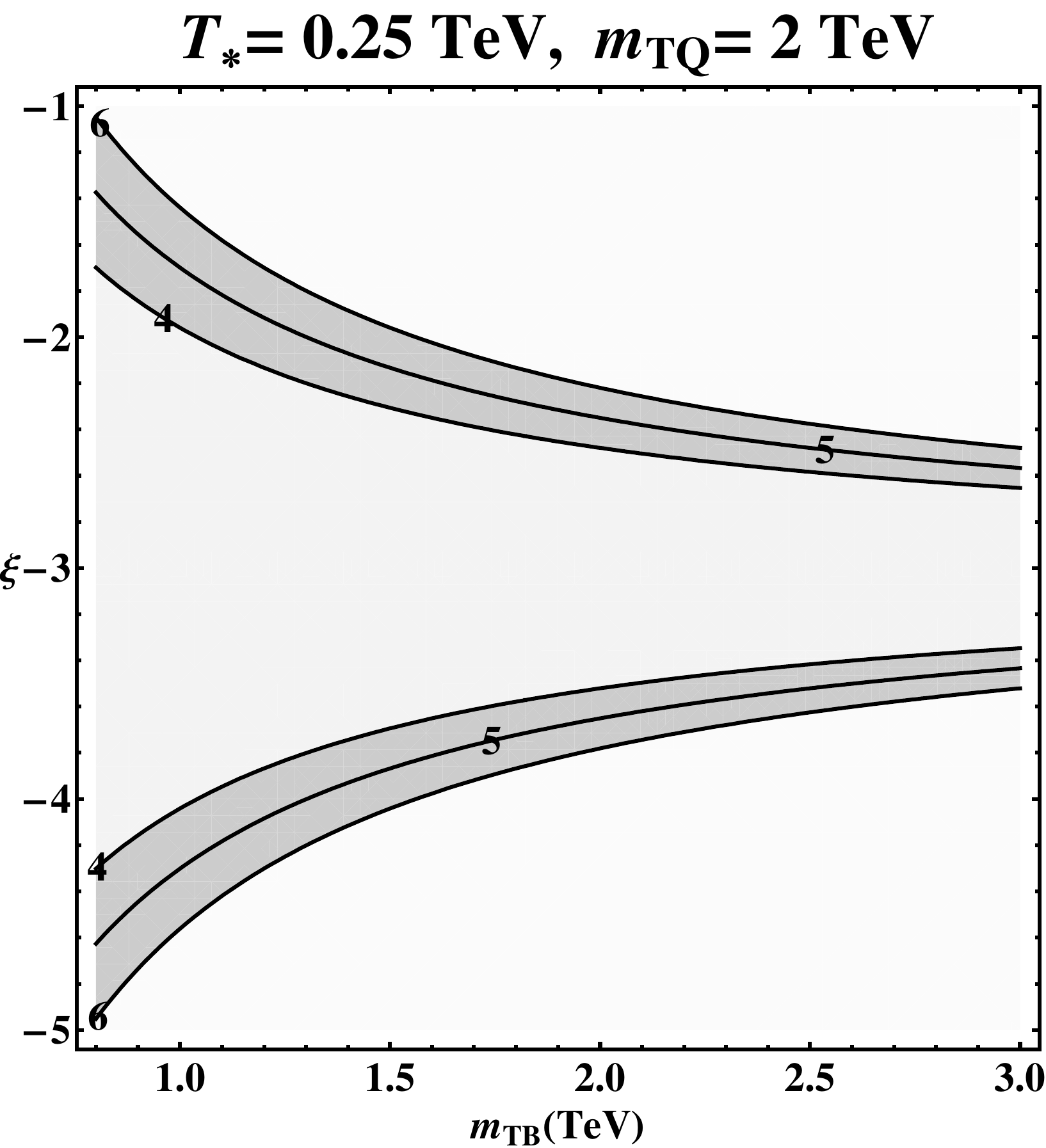}}
\caption{Contour plot diagram in the $\xi$ and $m_{TB}$ parameter space representing different values assumed by  the ratio $\Omega_{TB}/\Omega_B$.  Within our approximations the regions depend on the ratio $T_{\ast}/m_{TQ}$. $T_{\ast}$ is the temperature below which the processes violating the baryon, technibaryon and lepton numbers cease to be relevant and $m_{TQ}$ the dynamical mass of the techniquarks. The plots correspond to eight distinct values of this ratio. }\label{fig:DM}
\end{figure}

In Fig. \ref{fig:DM} we show the contour plot diagram in the $\xi - m_{TB}$ plane representing different values assumed by  the ratio $\Omega_{TB}/\Omega_B$.  Within our approximations the regions depend on the ratio $T_{\ast}/m_{TQ}$, where $T_{\ast}$ is the temperature below which the processes violating the baryon, technibaryon and lepton numbers cease to be relevant and $m_{TQ}$ is the dynamical mass of the techniquarks. The plots correspond to eight distinct values of this ratio. The two regions  having  $4\leq \Omega_{TB}/\Omega_B \leq 6$ are in dark gray. In between these two regions the ratio diminishes while in the upper and lower part the ratio increases.

What is interesting is that, differently from the case in which the technibaryon acquires mass only due to technicolor interactions, one achieves the desired phenomenological ratio of DM to baryon matter with a light technibaryon mass with respect to the weak interaction scale. In fact the mass can be even lower than $100$~{\rm GeV}. This DM candidate can be produced at the Large Hadron Collider experiment.

 To provide a simple estimate for the TIMP-nucleus cross section useful for the CDMS searches we adopt the model
computations provided in \cite{Bagnasco:1993st}. We note first that the TIMP does not interact directly with the SM. The dominant scalar TIMP - nucleus cross section is suppressed by at least four powers of the technicolor dynamical scale.

Following \cite{Bagnasco:1993st} the total number of counts $R$ per unit detector mass $m$ and nuclear recoil kinetic energy $E_R$ in the lab frame is
\begin{eqnarray}
\frac{dR}{dm\ dE_R} \simeq 1.38\times 10^{-4} | F_c(E_R) |^2\ \Lambda_{\text{TeV}}^{-4}\  M^{-1}_{\text{TeV}}\ \rho_{0.3}\ V^{-1}_{220}\ \left( \text{kg}\ \text{keV}\ \text{day} \right)^{-1}
\end{eqnarray}
where $F_c(E_R)$ is the scalar nuclear form factor which takes into account the finite size effects. In the expression above $\Lambda_{\text{TeV}} = \Lambda_{TC}/\text{TeV}$, $M_{\text{TeV}} = m_{TB}/\text{TeV}$, $\rho_{0.3} = \rho/\left( 0.3\ \text{GeV}\ \text{cm}^{-3} \right)$, $V_{220} = \frac{V_0}{\left( 220\ \text{km}\ \text{s}^{-1} \right)}$ with $\rho$ and $V_0$ being the technibaryon density and a suitably weighted average velocity respectively.

\begin{figure}[t]
{\includegraphics[height=6.5cm,width=9cm]{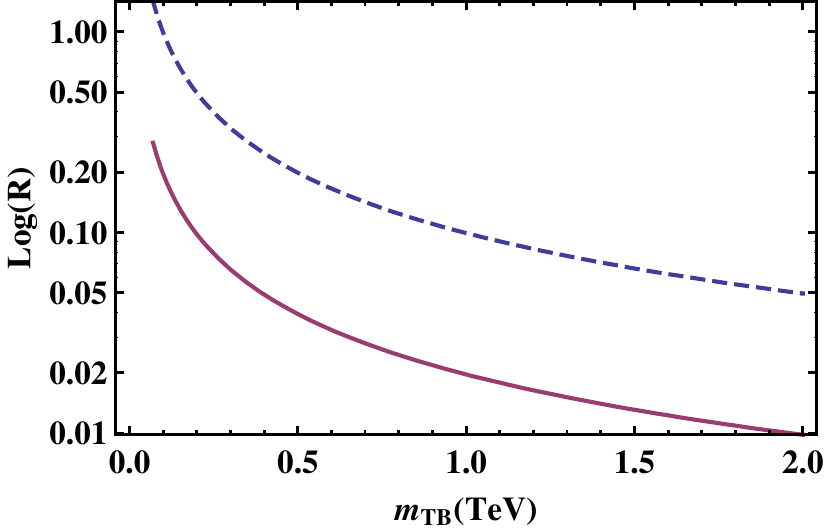}\hspace{0.8cm}}
\caption{The expected number of counts in a Germanium detector for an effective exposure of $121.3\ (\text{kg}\ \text{day})$ and recoil energies in the range $5-100\ \text{keV}$. The dashed curve corresponds to $\Lambda_{TC} = 2\ \text{TeV}$ while the solid curve corresponds to $\Lambda_{TC} = 3\ \text{TeV}$.}\label{fig:counts}
\end{figure}

To compare our predictions with the CDMS results we plot in Fig. \ref{fig:counts} the total number of expected counts for an effective exposure of $121.3\ (\text{kg}\ \text{day})$ and recoil energies in the range $5-100\ \text{keV}$. The dashed curve corresponds to $\Lambda_{TC} = 2\ \text{TeV}$ while the solid curve corresponds to $\Lambda_{TC} = 3\ \text{TeV}$.

Our TIMP is a template for a more general class of models according to which the lightest one is neutral under the SM interactions. Models belonging to this class are, for example, partially gauged technicolor.

\section{Summary and Outlook}

We proposed a technicolor model with technifermion matter transforming according to two distinct representations of the underlying technicolor gauge group. The model features simultaneously the smallest possible value of the naive $S$ parameter and the smallest possible number of technifermions. The chiral dynamics is intriguing and very rich. After having classified the relevant low energy composite spectrum we have constructed the associated effective Lagrangians. We introduced both the linear and non-linearly realized one.
The linearly realized one will permit us to study immediately the thermal properties of the chiral phase transition relevant for electroweak baryogenesis as done for the case of Minimal Walking Technicolor \cite{Cline:2008hr}. Due to the interplay between multiple nearby phase transitions \cite{Mocsy:2003qw,Sannino:2004ix} we expect novel phenomena of direct interest for cosmological applications. The linearly realized Lagrangian, once extended to contain also the spin one composite spectrum, will be of immediate interest for LHC phenomenology. The construction of the non-linear Lagrangian is interesting, instead, since it is exact in the limit of small momenta, at least untill the first resonance is encountered. It will also allow to neatly incorporate the non-abelian anomalies and the associated topological terms
\cite{Witten:1983tw,Witten:1983tx,Wess:1971yu,Kaymakcalan:1984bz,Kaymakcalan:1983qq,Duan:2000dy} as well as the study of the its solitonic excitations.

We have embedded, in a natural way, the SM interactions within the global symmetries of the underlying gauge theory. Several low-energy composite particles were found to be SM singlets. At least one of these  TIMPs has been recognized as a promising cold DM candidate. The novel TIMP can be sufficiently light, with respect to the technicolor dynamical scale, to be directly produced at the LHC and simultaneously constrained by the CDMS experiment.

\acknowledgments
The work of T.R is supported by a Marie Curie Early Stage Research Training Fellowship of the European Community's Sixth Framework Programme under contract number MEST-CT-2005-020238-EUROTHEPHY. The work of F.S. is supported by the Marie Curie Excellence Grant under contract MEXT-CT-2004-013510.

\appendix

\section{SU(4)$\times$SU(2)$\times$U(1) Generators} \label{appendixgenerators}
Here we construct the explicit realization of the generators of $SU(4)\times SU(2)\times U(1)$. We denote the fifteen generators of $SU(4)$ by $S_4^a$ and $X_4^i$ with $a=1,\ldots,10$ and $i=1,\ldots,5$. They can be represented as:
\begin{eqnarray}
S_4^a = \left( \begin{array}{cc}
\textbf{A} & \textbf{B} \\
\textbf{B}^{\dagger} & -\textbf{A}^T
\end{array} \right) \ , \qquad X_4^i = \left( \begin{array}{cc}
\textbf{C} & \textbf{D} \\
\textbf{D}^{\dagger} & \textbf{C}^T
\end{array} \right) \ ,
\end{eqnarray}
where $A$ is Hermitian, $C$ is Hermitian and traceless, $B$ is symmetric and $D$ is antisymmetric. The $S_4^a$ obey the relation $(S_4^a)^TE + ES_4^a = 0$ and are a representation of $Sp(4)$. They are explicitly given by:
\begin{eqnarray}
S_4^a &=& \frac{1}{2\sqrt{2}} \left( \begin{array}{cc}
\tau^a & \textbf{0} \\
\textbf{0} & -\tau^{aT}
\end{array} \right) \ , \qquad a=1,\ldots,4 \\
S_4^a &=& \frac{1}{2\sqrt{2}} \left( \begin{array}{cc}
\textbf{0} & \textbf{B}^a \\
\textbf{B}^{a\dagger} & \textbf{0}
\end{array} \right) \ , \qquad a=5,\ldots,10
\end{eqnarray}
where $\tau^{1,2,3}$ are the usual Pauli matrices, $\tau^4=\textbf{1}$ and:
\begin{eqnarray}
\begin{array}{ccc}
B^5 = \textbf{1} \ , & B^7=\tau^3 \ , & B^9=\tau^1 \ , \\
B^6 = i\textbf{1}\ , & B^8 = i\tau^3\ , & B^{10} = i\tau^1\ .
\end{array}
\end{eqnarray}

The remaining five generators are explicitly given by:
\begin{eqnarray}
X_4^i &=& \frac{1}{2\sqrt{2}} \left( \begin{array}{cc}
\tau^i & \textbf{0} \\
\textbf{0} & \tau^{iT}
\end{array} \right)\ , \qquad i=1,\ldots,3 \\
X_4^i &=& \frac{1}{2\sqrt{2}} \left( \begin{array}{cc}
\textbf{0} & \textbf{D}^i \\
\textbf{D}^{i\dagger} & \textbf{0}
\end{array} \right) \ , \qquad i=4,5
\end{eqnarray}
with:
\begin{eqnarray}
D^4= \tau^2 \ , \qquad D^5 = i\tau^2 \ .
\end{eqnarray}

The generators are normalized according to:

\begin{eqnarray}
\text{Tr} \left[ S_4^aS_4^b \right] = \frac{1}{2} \delta^{ab} \ , \qquad \text{Tr} \left[ X_4^iX_4^j \right] = \frac{1}{2} \delta^{ij} \ , \qquad \text{Tr}  \left[ S_4^aX_4^i \right] = 0 \ .
\end{eqnarray}

The generators of $SU(2)$ are similarly divided into the two that are broken $X_2^i =\frac{\tau^i}{2},\ i=1,2$ and the one that leaves the vacuum invariant $S_2^1 = \frac{\tau^3}{2}$.

For convenience we shall consider the nineteen generators of $SU(4)\times SU(2)\times U(1)$ as $6\times 6$ block diagonal matrices. They are denoted by $S^a,\ a=1,\ldots,11$ and $X^i,\ i=1,\ldots,8$. The eleven generators $S^a$ are a representation of the subgroup $Sp(4)\times SO(2)$ and are given by

\begin{eqnarray}
S^a &=& \left( \begin{array}{cc}
S_4^a & \\
 & 0_{2\times 2}
\end{array} \right) \ , \qquad a=1,\ldots,10 \\
S^{11} &=& \left( \begin{array}{cc}
0_{4\times 4} & \\
 & S_2^1
\end{array} \right) \ .
\end{eqnarray}
while the remaining eight generators are given explicitly by
\begin{eqnarray}
X^i &=& \left( \begin{array}{cc}
X_4^i & \\
 & 0_{2\times 2}
\end{array} \right) \ , \qquad i=1,\ldots,5 \\
X^i &=& \left( \begin{array}{cc}
0_{4\times 4} & \\
 & X_2^{i-5}
\end{array} \right) \ , \qquad i=6,7 \\
X^8 &=& \frac{1}{3}\text{diag}(-1,-1,-1,-1,\frac{1}{2}, \frac{1}{2})
\end{eqnarray}

They are normalized according to:

\begin{eqnarray}
\text{Tr} \left[ S^aS^b \right] = \frac{1}{2}\delta^{ab} \ , \qquad \text{Tr} \left[ X^i X^j \right] = \frac{1}{2} \delta^{ij} \ , \qquad  \text{Tr} \left[ S^a X^i \right] = 0 \ .
\end{eqnarray}

\section{Dark Matter Computations} \label{darkmatter}

We follow the notation and analysis of \cite{Harvey:1990qw} and \cite{Gudnason:2006yj},
and denote the chemical potentials of the SM particles by
\begin{equation}
\begin{array}{rclcrcl}
  \mu_W &\ \ \text{for}\ \ & W^- & ,~~~~~~~~~~~~ & \mu_{dL} &\ \ \text{for}\ \ & d_L,s_L,b_L \ , \\
  \mu_0 &\ \ \text{for}\ \ & \phi^0 & ,~~~~~~~~~~~~ & \mu_{dR} &\ \ \text{for}\ \ & d_R,s_R,b_R \ , \\
  \mu_- &\ \ \text{for}\ \ & \phi^- & ,~~~~~~~~~~~~ & \mu_{iL} &\ \ \text{for}\ \ & e_L,\mu_L,\tau_L \ , \\
  \mu_{uL} &\ \ \text{for}\ \ & u_L,c_L,t_L & ,~~~~~~~~~~~~ & \mu_{iR} &\ \ \text{for}\ \ & e_R,\mu_R,\tau_R \ , \\
  \mu_{uR} &\ \ \text{for}\ \ & u_R,c_R,t_R & ,~~~~~~~~~~~~ & \mu_{\nu iR} &\ \ \text{for}\ \ & \nu_{eR},\nu_{\mu R},\nu_{\tau R} \ , \\
  \mu_{\nu iL} &\ \ \text{for}\ \ & \nu_{eL},\nu_{\mu L}, \nu_{\tau L} & ,~~~~~~~~~~~~
\end{array}
\end{equation}
while the chemical potentials of the new particles are denoted by
\begin{equation}
\begin{array}{rclcrcl}
  \mu_{UL} &\ \ \text{for}\ \ & U_L & ,~~~~~~~~~~~~ & \mu_{DL} &\ \ \text{for}\ \ & D_L \ , \\
  \mu_{UR} &\ \ \text{for}\ \ & U_R & ,~~~~~~~~~~~~ & \mu_{DR} &\ \ \text{for}\ \ & D_R \ , \\
\end{array}
\end{equation}
The two components of the SM-type Higgs doublet are denoted as $\phi_-$ and $\phi_0$. These translate in our notation to $\phi_-= \Pi^-$ and $\phi_0 = {\sigma_4 - i \Pi^0}$. We have assigned the same chemical potential for the SM triplet $u,c,t$ and $d,s,b$ respectively and minimally coupled the composite Higgs to the SM fermions assuming, for the Yukawa sector, the existence of a working ETC dynamics.

Thermal equilibrium in the electroweak interactions implies the following relations among the chemical potentials of the SM particles
\begin{equation}
\begin{array}{rclcrcl}\label{SMrelations}
  \mu_{W} & = & \mu_- +\mu_0 & ,~~~~~~~~~~~~ & W^- & \leftrightarrow & \phi^- + \phi^0  \ , \\
  \mu_{dL} & = & \mu_{uL} +\mu_{W} & ,~~~~~~~~~~~~ & W^- & \leftrightarrow & \bar{u}_L + d_L  \ , \\
  \mu_{iL} & = & \mu_{\nu iL} +\mu_{W} & ,~~~~~~~~~~~~ & W^- & \leftrightarrow & \bar{\nu}_{iL} + e_{iL}  \ , \\
  \mu_{\nu iR} & = & \mu_{\nu iL} +\mu_{0} & ,~~~~~~~~~~~~ & \phi^0 & \leftrightarrow & \bar{\nu}_{iL} + \nu_{iR}  \ , \\
  \mu_{uR} & = & \mu_{0} +\mu_{uL} & ,~~~~~~~~~~~~ & \phi^0 & \leftrightarrow & \bar{u}_{L} + u_{R}  \ , \\
  \mu_{dR} & = & -\mu_{0} +\mu_{W} + \mu_{uL} & ,~~~~~~~~~~~~ & \phi^0 & \leftrightarrow & d_L + \bar{d}_R  \ , \\
  \mu_{iR} & = & -\mu_{0} +\mu_{W} + \mu_{\nu iL} & ,~~~~~~~~~~~~ & \phi^0 & \leftrightarrow & e_{iL} + \bar{e}_{iR}  \ , \\
\end{array}
\end{equation}
and the following relations among the chemical potentials of the techniquarks
\begin{equation}\label{TCrelations}
\begin{array}{rclcrcl}
  \mu_{DL} & = & \mu_{UL} +\mu_{W} & ,~~~~~~~~~~~~ & W^- & \leftrightarrow & \bar{U}_L + D_L  \ , \\
   \mu_{UR} & = & \mu_{0} +\mu_{UL} & ,~~~~~~~~~~~~ & \phi^0 & \leftrightarrow & \bar{U}_{L} + D_{R}  \ , \\
   \mu_{DR} & = & -\mu_{0} +\mu_{W} + \mu_{UL} & ,~~~~~~~~~~~~ & \phi^0 & \leftrightarrow & D_L + \bar{D}_R  \ , \\
\end{array}
\end{equation}
The thermodynamical analysis is most transparent when using directly the underlying technicolor degrees of freedom. At a given temperature $T$ and chemical potential $\mu$ the number density $n_+$ ($n_-$) of particles (antiparticles) is given by
\begin{eqnarray}
n_{\pm} = m\int \frac{d^3k}{\left(2\pi \right)^3} \frac{1}{z^{\mp 1} e^{E\beta} - \eta}
\end{eqnarray}

Here $m$ is the multiplicity of the degrees of freedom, $\beta =1/T$, $z=e^{\mu \beta}$ is the fugacity, $E^2=m^2 + \vec{k}^2$ is the energy and $\eta$ equals $1$ and $-1$ for bosons and fermions respectively.

At the freeze-out temperature $T^*$, where the violating processes cease to be efficient, we have $\mu/T^* \ll 1$ and we therefore find that the difference between the number densities of particles and their corresponding antiparticles is
\begin{eqnarray}
n = n_+ - n_- = mT^{*3}\cdot \frac{\mu}{T^*}\cdot \frac{\sigma\left( \frac{m}{T^*} \right)}{6}
\end{eqnarray}
where we have defined the statistical function $\sigma$ as
\begin{eqnarray}
\sigma(z) = \left\{ \begin{array}{rl}
\frac{6}{4\pi^2} \int_{0}^{\infty} dx\ x^2\cosh^{-2} \left( \frac{1}{2} \sqrt{x^2 + z^2 } \right) &\qquad  \text{for fermions} \ , \\
\frac{6}{4\pi^2} \int_{0}^{\infty} dx\ x^2\sinh^{-2} \left( \frac{1}{2} \sqrt{x^2 + z^2 } \right) &\qquad \text{for bosons} \ .
\end{array} \right.
\end{eqnarray}

We have conveniently normalized the statistical function such that it assumes the value $1$ ($2$) for massless fermions (bosons). When computing the relic density we are only interested in the ratio of number densities. Hence we appropriately normalize the net baryon number density as:
\begin{eqnarray}
B = \frac{6}{mT^{*2}} \left(n_B - n_{\bar{B}}\right)
\end{eqnarray}

A similar normalization is chosen for the lepton and technibaryon number densities. Having set the notation the overall electric charge is
\begin{eqnarray}
Q &=& \frac{2}{3}\cdot 3 \left( 2+ \sigma_t \right) \left( \mu_{uL} + \mu_{uR} \right) - \frac{1}{3} \cdot 3 \cdot 3 \left( \mu_{dL} + \mu_{dR} \right) - \sum_i \left(\mu_{iL} + \mu_{iR} \right)  \nonumber \\
&&  - 2\cdot 2 \mu_{W} - 2\mu_- + \frac{1}{2}\cdot 2 \sigma_{U} \left( \mu_{UL} + \mu_{UR} \right) -  \frac{1}{2}\cdot 2 \sigma_{D} \left( \mu_{DL} + \mu_{DR} \right) \nonumber \\
&=& 2 \left( \sigma_{U} - \sigma_{D} \right) \mu_{UL} + 2\left( 1+2\sigma_t \right) \mu_{uL} - 2 \left( 9 + \sigma_{D} \right) \mu_W  \nonumber \\
&& -2 \mu + \left( 12+ 2\sigma_t + \sigma_{U} + \sigma_{D} \right)\mu_0
\end{eqnarray}
with $\mu = \sum_i \mu_{\nu i L}$ while the overall weak isospin charge is
\begin{eqnarray}
Q_3 &=& \frac{1}{2} \cdot 3\cdot \left(2 + \sigma_t \right) \mu_{uL} -\frac{1}{2} \cdot 3\cdot 3 \mu_{dL} + \frac{1}{2} \sum_i \left( \mu_{\nu iL} - \mu_{iL} \right) - 4 \mu_W \nonumber \\
&& - \left( \mu_0 + \mu_- \right) + \frac{1}{2}\cdot 2 \sigma_{U} \mu_{UL} - \frac{1}{2}\cdot 2 \sigma_{D} \mu_{DL} \nonumber \\
&=& \frac{3}{2} \left( \sigma_t-1 \right) \mu_{uL} - \left( 11 + \sigma_{D} \right) \mu_{W} + \left( \sigma_{U} - \sigma_{D} \right) \mu_{UL} \ .
\end{eqnarray}

Here we have used the relations \ref{SMrelations} and \ref{TCrelations}. Relation between chemical potentials coming from baryon number violating processes:
\begin{eqnarray}
0 &=& \mu_{UL} + \mu_{DL} +3 \left( \mu_{uL} + 2\mu_{dL} \right) + \mu \\
 &=& 2 \mu_{UL} + 9 \mu_{uL} + 7 \mu_W + \mu \ .
\end{eqnarray}

Finally we note that the baryon number $B$, lepton number $L$ and technibaryon number $TB$ can be expressed as
\begin{eqnarray}
B &=& \left( 10 + 2 \sigma_t \right) \mu_{uL} + 6 \mu_W + \left( \sigma_t - 1 \right) \mu_0 \\
L &=& 6 \mu_W + 4\mu \\
TB &=& \frac{1}{2\sqrt{2}} \cdot 2 \left[ \sigma_{U} \left( \mu_{UL} + \mu_{UR} \right) + \sigma_{D} \left( \mu_{DL} + \mu_{DR} \right) \right] \nonumber \\
&=& \frac{1}{\sqrt{2}} \left[ 2 \left( \sigma_{U} + \sigma_{D} \right)\mu_{UL} + 2 \sigma_{D} \mu_W + \left( \sigma_{U} - \sigma_{D} \right) \mu_0 \right]
\end{eqnarray}

\subsection{2nd Order Phase Transition}

Here we have the following conditions: $Q=0$ and $\mu_0=0$. In the approximation where the up and down techniquarks have equal masses we find, using the relations above, that the technibaryon number over the baryon number can be written as:
\begin{eqnarray}
- \frac{\sqrt{2}\cdot TB}{B} = \frac{\sigma}{81+10\sigma + \left( 27 +2 \sigma \right) \sigma_t} \left[ 18\left( 8 + \sigma_t + \sigma \right) + \left( 5+ \sigma_t \right) \left( 9 + \sigma \right) \xi \right]
\end{eqnarray}
where $\sigma = \sigma_{U} = \sigma_{D}$ and $\xi=L/B$.

\subsection{1st Order Phase Transition}

For the first order phase transition we impose the following two conditions: $Q=0$ and $Q_3=0$. We then find that:
\begin{eqnarray}
- \frac{\sqrt{2}\cdot TB}{B} &=& \frac{\sigma}{2513+654\sigma + 40\sigma^2 + 2 \left( 551+ 102\sigma + 4\sigma^2 \right)\sigma_t + \left( 81+ 6\sigma \right) \sigma_t^2} \nonumber \\
&& \times \bigg[ 18\left( 246 +65\sigma + 4\sigma^2 + \left( 59+7\sigma \right) \sigma_t + 3\sigma_t^2 \right) \nonumber  \\
&& + \left( 1441 + 345\sigma + 20\sigma^2 + 4 \left( 95 +21\sigma + \sigma^2 \right) \sigma_t +3\left( 9+\sigma \right)\sigma_t^2 \right) \xi \bigg]
\end{eqnarray}

In the approximation where the top quark is also considered massless the technibaryon number over the baryon number is the same for both the 1st and 2nd order phase transition
\begin{eqnarray}
- \frac{\sqrt{2}\cdot TB}{B} &=& \frac{\sigma}{2} \left( 3 + \xi \right) \ .
\end{eqnarray}

\end{document}